\documentclass{jkas}

\usepackage[caption=false]{subfig}
\usepackage{graphicx}
\usepackage{epstopdf}
\usepackage{amsmath}

\usepackage{color}
\usepackage[]{units}
\usepackage[]{nth}

\usepackage{xcolor}

\def\year{2018} % publication year
\def\month{} % publication month
\def\volume{00} % journal volume
\def\issue{0} % journal issue
\def\beginpage{1} % first page of article
\def\endpage{99} % last page of article
\def\received{2018} % date paper was received by JKAS
\def\accepted{2018} % date of acceptance
\date{Received \received; accepted \accepted}

\title{New Photometric Pipeline to Explore Temporal and Spatial Variability with KMTNet DEEP-South Observations}

\author[1,2,3]{Seo-Won Chang}
\author[3]{Yong-Ik Byun}
\author[4]{Min-Su Shin}
\author[3]{Hahn Yi}
\author[4]{Myung-Jin Kim}
\author[4]{Hong-Kyu Moon}
\author[4]{Young-Jun Choi}
\author[4,5]{Sang-Mok Cha}
\author[4,5]{Yongseok Lee}

\affil[1]{Research School of Astronomy and Astrophysics, The Australian National University, Canberra, ACT 2611, Australia \email{seowon.chang@anu.edu.au}}
\affil[2]{ARC Centre of Excellence for All-sky Astrophysics (CAASTRO)}
\affil[3]{Department of Astronomy and University Observatory, Yonsei University, Seoul 03722, Republic of Korea}
\affil[4]{Korea Astronomy and Space Science Institute, Daejeon 34055, Republic of Korea}
\affil[5]{School of Space Research, Kyung Hee University, Gyeonggi 17104, Republic of Korea}

\def\corrauthor{
Seo-Won Chang
}

\def\runningauthor{
Chang et al.
}

\def\runningtitle{
Exploring variable objects with DEEP-South observations
}

\def\keywords{
methods: data analysis --- techniques: photometric --- stars: variables: general --- asteroids: general
}

\def\abstracttext{
The DEEP-South photometric census of small Solar System bodies produces massive time-series data of variable, transient or moving objects as a by-product. To fully investigate unexplored variable phenomena, we present an application of multi-aperture photometry and FastBit indexing techniques for  faster access to a portion of the DEEP-South year-one data. Our new pipeline is designed to perform automated point source detection, robust high-precision photometry and calibration of non-crowded fields which have overlap with previously surveyed areas. In this paper, we show some examples of catalog-based variability searches to find new variable stars and to recover targeted asteroids. We discover 21 new periodic variables with period ranging between 0.1 and 31 days, including four eclipsing binary systems (detached, over-contact, and ellipsoidal variables), one white dwarf/M dwarf pair candidate, and rotating variable stars. We also recover astrometry ($<\pm$1--2 arcsec level accuracy) and photometry of two targeted near-earth asteroids, 2006 DZ169 and 1996 SK, along with the small- ($\sim$0.12 mag) and relatively large-amplitude ($\sim$0.5 mag) variations of their dominant rotational signals in $R$-band.}

\begin{document}
\jkashead 
% Introduction
\section{Introduction\label{sec:intro}}
The Deep Ecliptic Patrol of the Southern Sky (DEEP-South: \citealt{moon2016}) is a dedicated photometric study to physically characterize small bodies in our Solar System, as one of the secondary science projects of Korea Microlensing Telescope Network (KMTNet; \citealt{kim2016a}). The DEEP-South employs a network of three identical 1.6-m telescopes located in Chile (CTIO), South Africa (SAAO) and Australia (SSO), allowing 24-hour monitoring of asteroids and comets. Light curves with $BVRI$-band colors have been acquired for more than two hundred Near-Earth Asteroids (NEAs) since late 2015. The major efforts to discover NEAs have been historically concentrated on planetary defense to ensure Earth's safety from asteroid impacts. Several NEA discovery projects, such as the Catalina Sky Survey (CSS, \citealt{lar2003}) and Panoramic Survey Telescope and Rapid Response System (Pan-STARRS, \citealt{kai2002}),  have cataloged of NEAs more than 90 percent of the estimated population larger than 140 meters in regular scans of the sky. Comparing with other NEA surveys, the advantage of the DEEP-South is the round-the-clock operation capability which is essential for either precision astrometry or time-series photometry with high temporal resolution.

 As in the cases of other asteroid surveys (e.g., LONEOS: \citealt{bow1995};  LINEAR: \citealt{sto2000}; CSS: \citealt{lar2003}), long-time baseline observations have the potential to advance our knowledge of variable and transient phenomena over different time-scales from RR Lyraes \citep{mic2008, ses2013, tor2015} to periodic variables \citep{pal2013, dra2014} to AGNs \citep{rua2012}. These time-domain data sets also increase our  understanding of stellar structure and evolution: pulsating stars (e.g.,  $\delta$ Scuti, RR  Lyrae, Cepheid, and Mira variables) are important probes of the internal structure in investigating different excitation mechanisms of oscillations in stars, eclipsing binary systems allow us to determine the masses of both stars in a direct way, and Type Ia supernovae are used as probes of cosmology. These new outcomes emphasize importance of accurate and homogeneous photometric measurements and calibrations in the entire survey data, in terms of further synergy with upcoming data releases by ongoing and future time-domain surveys. 

Since the standard software packages of the DEEP-South are designed for differential photometry of targeted moving objects \citep{yim2016}, we have implemented an updated version of the source detection and time-series photometry pipeline to recover point sources which are not extracted in the original pipeline. First, the new pipeline is designed to conduct robust high-precision photometry and calibration of non-crowded fields with a varying point spread function (PSF). The PSF varies across images and changes with time due to focus, pointing jitter, optical distortion or atmospheric conditions, particularly in cases of wide field-of-view (FoV). Due to the large FoV ($\sim$4 deg$^{2}$) of the KMTNet mosaic CCD images, we expect to see deformation of the image PSF as well as spatial variations in the pixel scale\footnote{\url{http://kmtnet.kasi.re.kr/kmtnet-eng/astrometric-calibration-for-kmtnet-data/}}. To mitigate this problem, we perform multi-aperture photometry in determining the optimal aperture for each point source at every epoch and correcting position-dependent variations in the PSF shape across the mosaic. In addition to this, the pipeline is designed to perform forced photometry, thus giving us the possibility to extract consistent measurement of explicit sources in every frame with respect to its reference (deep co-added) frame. This approach can reduce false detection rate at low flux levels.

Further, this paper addresses a multi-step calibration issue to tie all DEEP-South data coming from the three different telescopes to a consistent photometric system. As in the case of LINEAR and CSS surveys, standard stars are not available in every survey field for precision photometry, specially because of varying observation conditions. Due to the observing strategy of the program, however, we can determine relative calibration parameters (e.g., relative overall offset) by using multiply observed stars in any overlapping fields. Since some of these stars have been measured in other all-sky multi-color surveys, we can transform our Johnson-Cousins filter system (only $BVRI$) to other photometric systems properly taking into account color terms. Similar calibration approaches have been successfully used in other asteroid surveys \citep{dra2009,ses2011}.  

Lastly, we have implemented a database management system for handling very large source catalogs. The source database is a key part of public data release enabling future scientific investigations. The data from most asteroid surveys are easily accessible in the form of an SQL-based relational database with an interactive web-based interface, containing all the epoch-based source photometry and metadata for more than tens to hundreds of million objects (e.g., LINEARdb: \citealt{ses2011}; Pan-STARRS1: \citealt{fle16}; ATLAS-VAR: \citealt{hei18}). For convenience, all these systems provide internally linked tables by using a unique label for every object that was obtained by grouping individual source detections at a given matching radius. We can significantly increase speed of query execution for already linked objects in this approach. However, it does not allow us to investigate possible transients, variables or moving objects within the database itself in a flexible way. In order to accelerate data access and reduce query response time, we adopt an efficient data indexing technique called FastBit \citep{wu2009} that stores data in a column-oriented manner unlike traditional relational databases. In this paper, we focus on development and applications of catalog-based searches for variability in stars with fixed coordinates using this database. In the second paper of this series, we will present additional application example to search and identify moving objects with motion vectors. 

This paper is organized as follows. In Section 2 we describe the DEEP-South survey and experimental time-series data. In Section 3 we outline the new photometry pipeline, calibration issues and the FastBit database system. In Section 4 we discuss a few example applications exploring temporal and spatial variability, especially periodic variable stars and targeted asteroids.  We conclude with a view to utilize the pipeline for massive variability searches with a full set of the DEEP-South survey data.

\begin{figure*}[!ht]
\centering
\includegraphics[width=140mm]{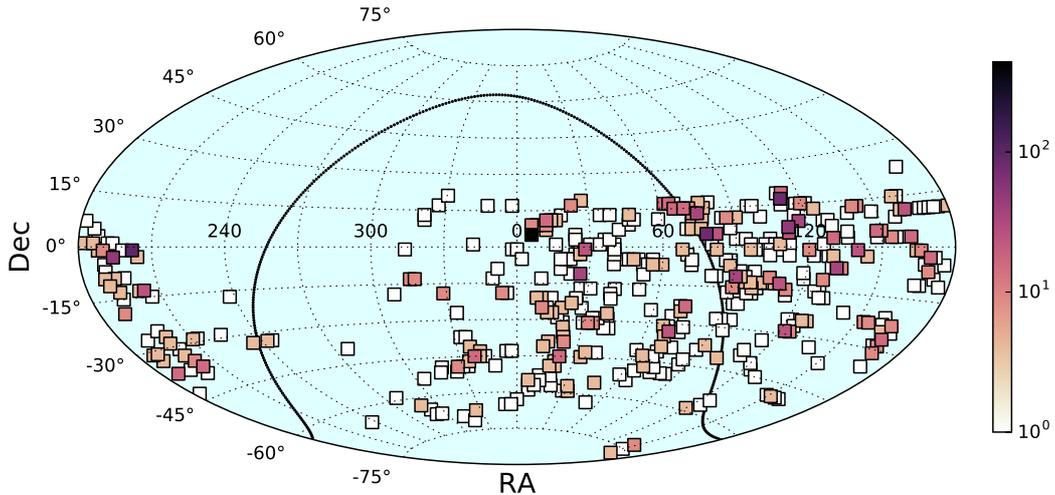}
\caption{Number of epochs per target field in the DEEP-South year-one data obtained from the three KMTNet telescopes. The values are color-coded by the bar on the right (from 1 to 510 visits). The solid line shows the Galactic plane in equatorial coordinates using the Aitoff projection. The location of the selected field, N02007-OC, is recognized by its darkest color in the center.\label{fig:jkasfig1}}
\end{figure*}

\begin{table*}
\caption{Summary of $R$-band time-series observations in the N02007-OC field\label{tab:jkastable1}}
\centering
\begin{tabular}{ccccc}
\toprule
Site        & Date & FWHM$^{\rm a}$ (arcsec) & Airmass & \# of Mosaic pointing \\
\midrule
CTIO & 2015-08-14 & 2.35$\pm$0.28 & 1.23--1.37 & 13 \\
      & 2015-08-16 & 1.37$\pm$0.17 & 1.22--1.31 & 15 \\
      & 2015-08-18 & 1.56$\pm$0.25 & 1.22--1.48 & 33 \\
      & 2015-08-22 & 3.52$\pm$0.44 & 1.22--1.55 & 35 \\
      & 2015-08-24 & 1.30$\pm$0.24 & 1.22--1.61 & 28 \\
      & 2015-11-01 & 1.38$\pm$0.16 & 1.22--1.47 & 26 \\
      & 2015-11-02 & 1.92$\pm$0.30 & 1.22--1.53 & 26 \\
      & 2015-11-03 & 1.47$\pm$0.24 & 1.22--1.62 & 25 \\
      & 2015-11-05 & 1.54$\pm$0.20 & 1.22--1.59 & 23 \\ %\addlinespace
      \midrule
SAAO & 2015-08-08 & 2.05$\pm$0.42 & 1.26--1.49 & 26 \\
      & 2015-08-10 & 1.28$\pm$0.44 & 1.26--1.49 & 32 \\
      & 2015-08-12 & 2.72$\pm$0.21 & 1.28--1.38 & 9 \\
      & 2015-08-14 & 2.68$\pm$0.33 & 1.26--1.39 & 23 \\
      & 2015-08-24 & 2.91$\pm$0.38 & 1.26--1.78 & 31 \\
      & 2015-11-09 & 2.79$\pm$0.49 & 1.35--1.58 & 12 \\ %\addlinespace 
      \midrule
SSO & 2015-08-08 & 1.97$\pm$0.26 & 1.24--1.39 & 26 \\
      & 2015-08-10 & 2.38$\pm$0.33 & 1.24--1.46 & 29 \\
      & 2015-08-18 & 1.60$\pm$0.20 & 1.24--1.58 & 35 \\
      & 2015-08-20 & 1.40$\pm$0.26 & 1.24--1.61 & 32 \\
      & 2015-08-22 & 2.23$\pm$0.22 & 1.24--1.36 & 17 \\
      & 2015-11-07 & 3.02$\pm$0.52 & 1.31--1.62 & 14 \\ %\addlinespace
      \midrule
Total &                  &                              &                   &   510 \\     
\bottomrule
\end{tabular}
\tabnote{
$^{\rm a}$  Median over images on that date.
}
\end{table*}

% Section 2
\section{The DEEP-South Photometric Census for Asteroids and Comets\label{sec:dsproject}}
The DEEP-South observations were made with three 1.6-m telescopes that have prime focus during the off-season of the Galactic bulge monitoring campaign of the KMTNet main survey. Survey observations comprise 135 distinct full nights per year between 2015 and 2019. The KMTNet camera consists of four 9k by 9k CCDs having a FoV of 2$\times$2 degrees with a pixel scale of 0.4 arcsec per pixel \citep{kim2016a}. The CCD chips are arranged with vertical (north-south direction) and horizontal (east-west direction) gaps of about 373 and 184 arcsec, respectively. The cameras have a 30 second readout time for the full mosaic with a readout noise of 10 electrons. The DEEP-South survey uses four standard Johnson-Cousins $BVRI$ filters, but they used mostly $R$-band for time-series work. The quantum efficiency is 80--90\% across the 4000--9000$\AA$ range, with a peak at $R$-band ($\sim$90\%).  About 50--100 science and associated calibration frames are produced in every nightly run (total 65--260 GB of data products). All the raw images are transferred to the KMTNet data center located at Daejeon in Korea for pre-processing.

The survey is designed to address a number of scientific goals that use five different observation modes (see Table 1 in \citealt{moon2016} for details). Opposition Census (OC) is the most frequently used mode for targeted photometry of NEAs in the regions on the sky around opposition. At that location, their color or brightness can be measured accurately. Each OC run consists of a series of exposures of more than two target fields that are visible in the opposition region. They normally employ exposure times short enough to avoid reducing the detection sensitivity for a moving object which may appear slightly streaked. From these OC observations, we can expect to obtain physical properties (e.g., rotation period and color indices), and to discover unknown moving objects. They also devote a small fraction of time to conduct both ecliptic survey and target-of-opportunity follow-up observations.
 
% Section 2.1
\subsection{Experimental Data Sets\label{sec:data}}
For the purpose of developing a new photometric pipeline, we use the DEEP-South first-year data collected from late-July 2015 through mid-April 2016. Here we present results for the most frequently observed field, N02007-OC, centered at RA=00:24:00, DEC=+05:00:00 (Figure \ref{fig:jkasfig1}). The first letter in the field designation indicates a sign of Ecliptic latitude: N and S for + and -, respectively. The next two digits are Ecliptic latitude in degrees; the last three digits are Ecliptic longitude in degrees. This field has full overlap with the Sloan Digital Sky Survey (SDSS) which provides colors of objects detected in the $ugriz$ system. The observation logs are summarized in Table \ref{tab:jkastable1}. We conducted time-series observations of the N02007-OC field only in the $R$-filter to measure periodic variability induced by rotation of the two targeted asteroids (see Section 4.3). These moving objects were detected in 200-second exposures with sufficient signal-to-noise ratio down to ${R} \sim 24$. 

Since early science commissioning observations were undertaken, there have been many updates and changes in CCD readout electronics and software for telescope operation. As part of quality assurance of the year-one data, we check problematic images (e.g., poor telescope tracking, readout error, or bad seeing) separately for each of the four CCD chips in mosaic, i.e., on a chip-by-chip basis. We define the individual amplifier images as the basic unit of our photometry pipeline, called a \textit{frame}. After excluding 24 bad frames both automatically as well as manually through the pipeline (see Section 3), 16296 frames were used for this study. 

% Section 3
\section{Data processing pipeline\label{sec:pipeline}}
The raw mosaic images are pre-processed by the KMTNet data reduction pipeline, including overscan, bias and flat-fielding reduction. The KMTNet camera produces strong crosstalk signals in multi-segment amplifiers for bright or saturated sources. The level of these electronic ghosts varies from frame to frame, but these can be removed by measuring crosstalk coefficients among frames within a single CCD image (see \citealt{kim2016b} for details). 

The components of our processing pipeline are separated into three basic units based on the format of data. Here we briefly summarize main procedures applied to the pre-processed data.  \begin{enumerate}
\item MEF unit (18432$\times$18464 pixels): we obtain an accurate astrometric solution used SCAMP \citep{ber2006} after correcting for geometric field distortion that is well described by a polynomial with third order terms measured from the center of the mosaic. This distortion pattern appear to be stable over time at three different KMTNet sites. The astrometric solution was determined as follows: (i) we run SCAMP separately for each exposure, and then (ii) we run it again to find a final solution that is consistent across all input frames. This astrometric solution is verified against the 2MASS reference system \citep{skr2006}. The resultant internal and external astrometric accuracy on overlapping sources are less than 0.02$\pm$0.01 and 0.09$\pm$0.06 arcsec in both RA and Dec directions, respectively. In the future, we will use the second data release of Gaia as input to validate astrometric analysis. 

\item CCD unit (9216$\times$9232 pixels): we remove cosmic-ray hits or hot-pixels using the parallelized version of the L.A.Cosmic algorithm \citep{van2001}\footnote{\url{https://github.com/cmccully/lacosmicx}}.

\item AMP unit (1152$\times$9232 pixels): we adopt the strategy of treating all 32 frames separately in the photometry and calibration processes in order to reduce processing time.  
\end{enumerate}

% Section 3.1
\subsection{Frame Catalog Generation\label{sec:photframe}}
Source extraction is the first step in making a complete frame catalog for all point sources either in fixed or varying positions. We use the source detection algorithm implemented in the SExtractor \citep{ber1996} in which each source is regarded as as set of connected pixels that exceed threshold above the local background. In this work, all connected regions with more than 5 pixels above 2.5-$\sigma$ of the local background were extracted. We also chose the following SExtractor parameters to optimize point source detection as well as to properly take into account blended sources: \texttt{DETECT\textunderscore MINAREA}=5, \texttt{DETECT\textunderscore THRESH}=2.5, \texttt{DEBLEND\textunderscore NTHRESH}=64, and \texttt{DEBLEND\textunderscore MINCONT}=0.0001. The average FWHM of each frame was used as a prior input for \texttt{SEEING\textunderscore FWHM} parameter. This process also produces several measurement characteristics such as FWHM, elongation, extraction flags, and star/galaxy classifier.

\begin{figure}[!ht]
\centering
\includegraphics[width=80mm]{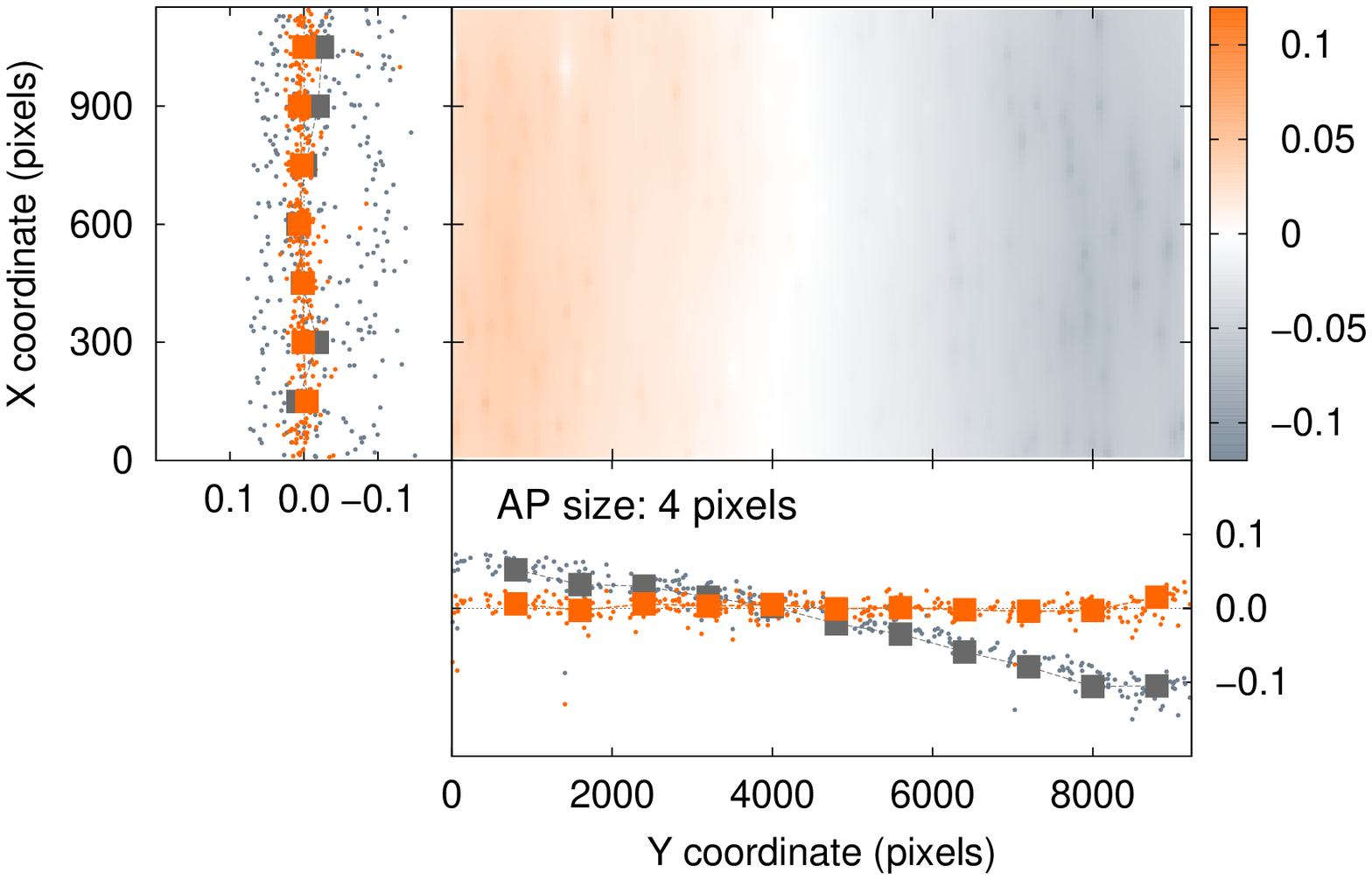}
\includegraphics[width=80mm]{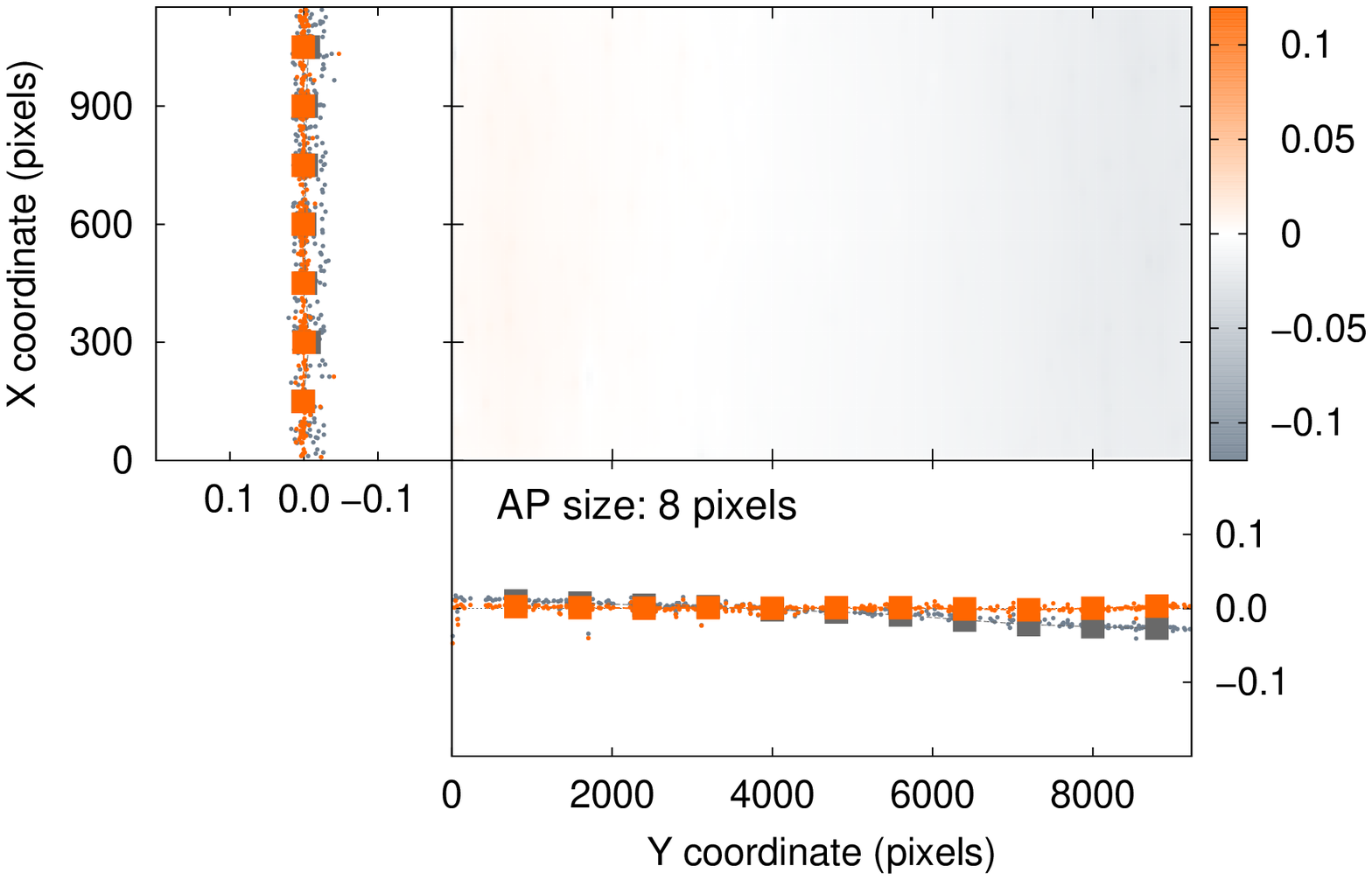}
\caption{Position- and aperture-dependent magnitude offsets (gray points) and that after the correction by a two-dimensional surface fitting (orange points) for an example frame. The dispersion of the magnitude offsets is much larger at smaller aperture sizes (top) than larger ones (bottom). The color maps show magnitude offsets in two-dimensional image frame. \label{fig:jkasfig2}} 
\end{figure}

We utilize the multi-aperture photometry algorithm \citep{cha2015} in order to: (i) determine the optimal aperture for each object at each epoch, (ii) use an empirical aperture-growth-curve method for flux correction, and (iii) add a new tag that isolates peculiar situations where photometry returns improper measurements. For each frame, we measure photometry in a series of circular apertures (up to 12-pixel radius) without changing the sky annulus, and we use the last aperture (corresponding to 12-pixel radii) as the reference for aperture corrections. We refer the reader to \citealt{cha2015} for details of the photometric performance of the pipeline. The position-dependent PSF variation and pixel-scale variation across fields for given focus result in up to 10\% changes in magnitude, depending on the selected aperture size for photometry. Figure \ref{fig:jkasfig2} shows an example of position- and aperture-dependent magnitude offsets, with the reminder that the coverage of x-axis ($\sim$0.13 deg) is about eight times smaller than that of y-axis ($\sim$1.02 deg). This mayby expected to artificially enlarge the apparent size of the magnitude offset along the y-axis. The magnitude offsets vary significantly for the case of photometry performed with small-sized apertures. The magnitude differences between the reference aperture and the smaller apertures are well described by two-dimensional polynomial forms, so we can consider any spatial variation in the aperture corrections across a single frame.

\begin{figure}[!ht]
\centering
\includegraphics[width=80mm]{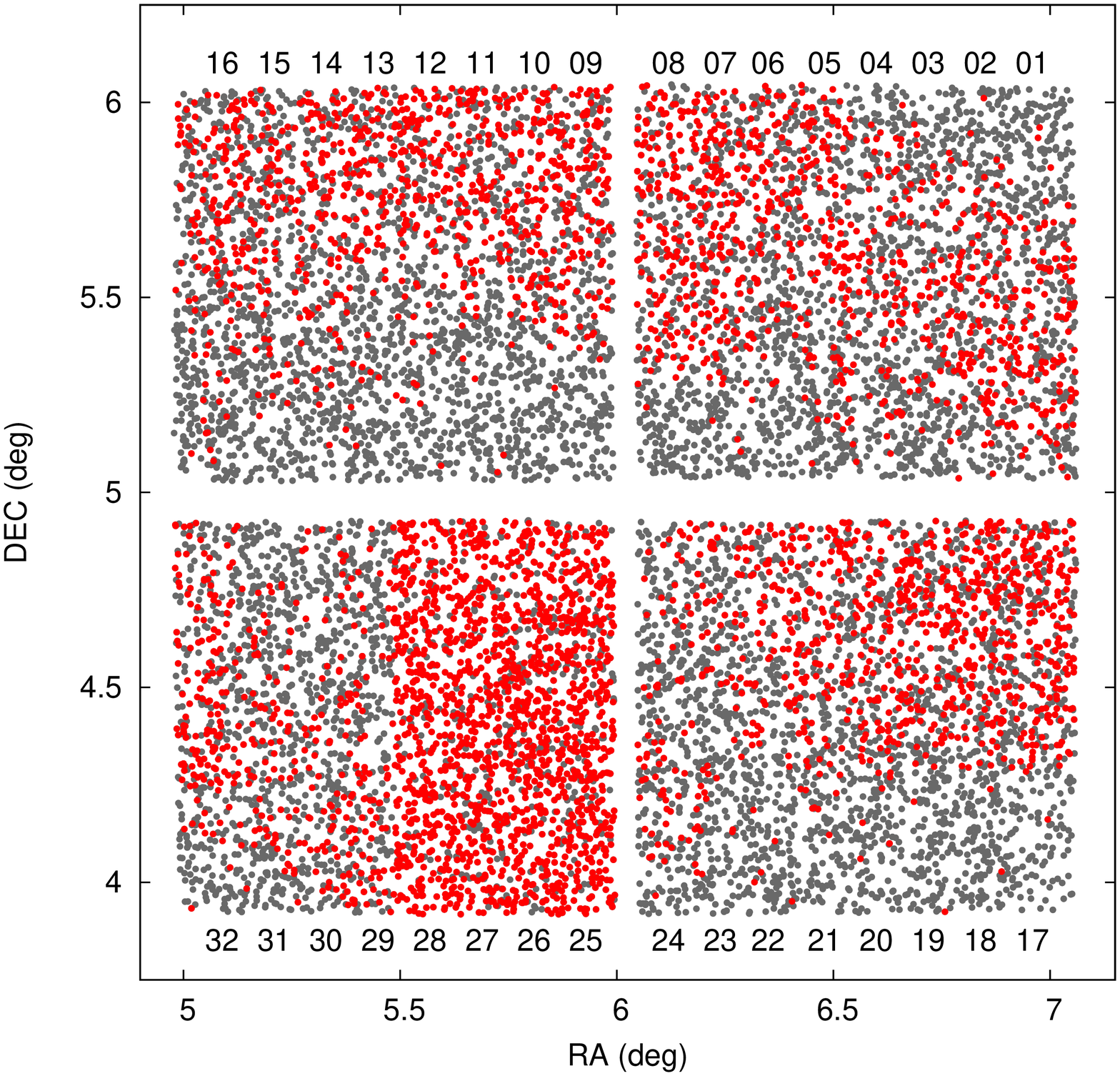}
\includegraphics[width=80mm]{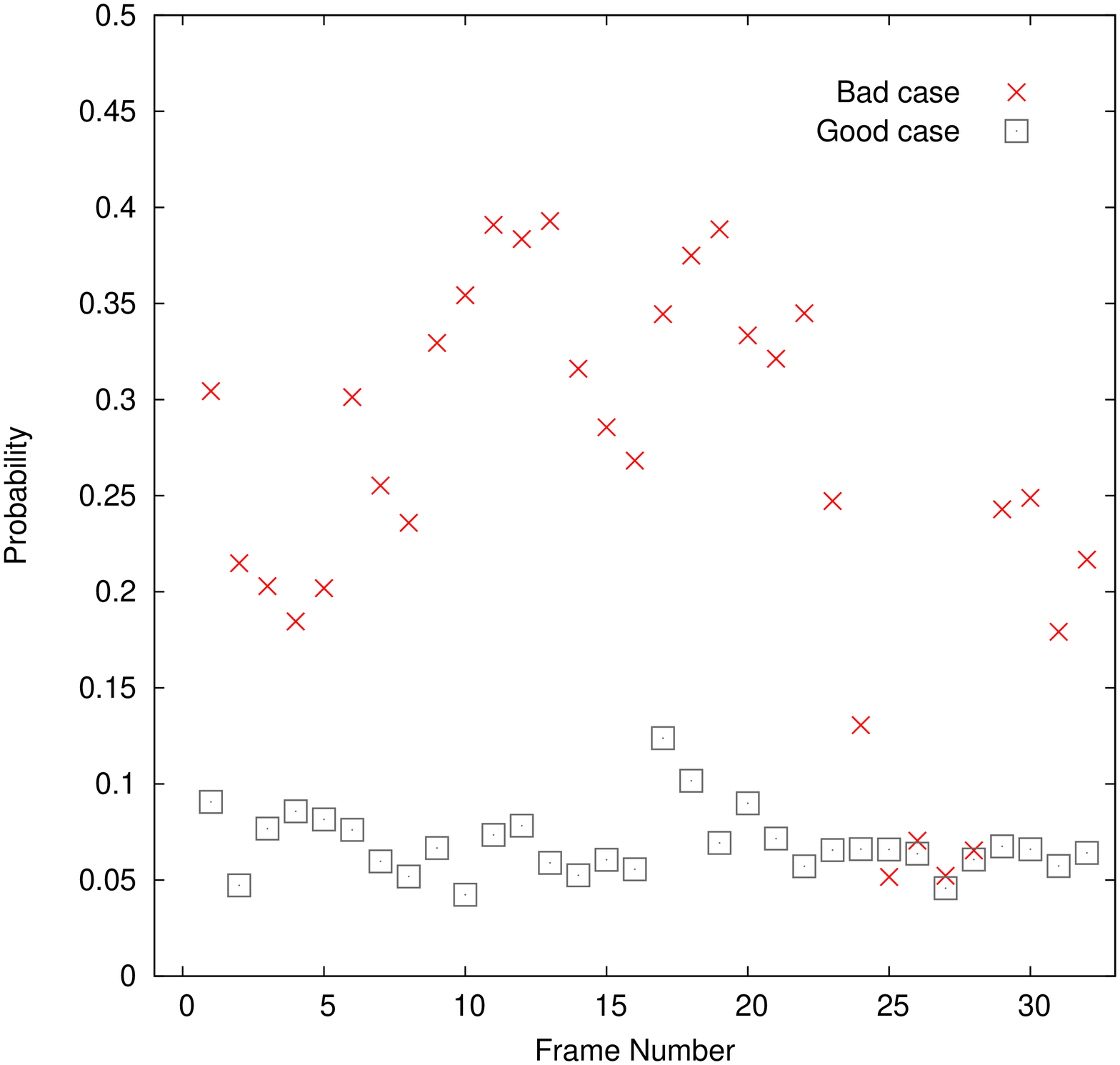}
\caption{Example of the spatial distribution of internal calibrators for good (gray points) and bad (red points) cases together with the calculated probability $P_{KS}$. In the top panel, each frame location is indicated by frame numbers.\label{fig:jkasfig3}} 
\end{figure}

Multiple visits to the same region of the sky enables us to achieve stable and uniform internal relative calibration in our instrumental photometric system. Bright, non-variable sources can be used as internal calibration stars, but it is necessary to check whether photometric quality of calibrators is sufficient and distributed homogeneously across the whole $(x, y)$ plane in a given frame. For the latter, we calculate an additional quality-ensuring cut characterizing two-dimensional distribution of these internal calibrators by using two-dimensional Kolmogorov-Smirnov (K-S) test (e.g., \citealt{pea1983}). The significance levels of flag statistic for the 2D K-S test can be summarized by the simple formula for the one-sample case (see also equation 14.7.1 in \citealt{pre1992}). Using this probability $P_{KS}$, we can figure out whether the source distribution is close to uniform across the frame. Figure \ref{fig:jkasfig3} shows the spatial distributions of selected calibration stars for good and bad cases, in which relatively large value of probability ($P_{KS}$ $>$ 0.15) indicates the presence of inhomogeneity.

\begin{figure}[!t]
\centering
\includegraphics[width=80mm]{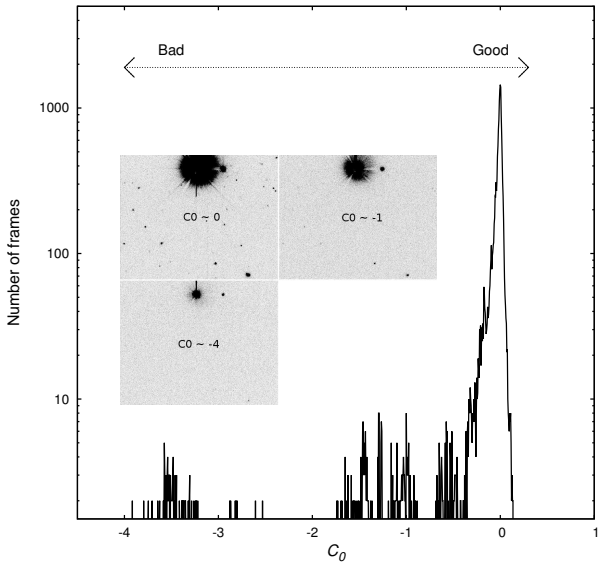}
\caption{Histogram of $C_{0}$ coefficients for all observed frames. The arrow indicates the range of photometric data quality observed under either perfectly ($C_{0}$ $\sim$ 0) or partially non-photometric conditions (-2 $<$ $C_{0}$ $<$ -1). All insert images are presented as an example of different quality cuts. \label{fig:jkasfig4}} 
\end{figure}

Following the suggestion of \citet{ive2007}, we also correct the spatial dependence of internal zeropoints around the photometric zeropoint separately for the 32 amplifiers in order to take into account atmospheric extinction gradients. We fit for linear gradients in both $x$ and $y$ coordinates simultaneously that can be expressed as:
\begin{equation}
\label{eq:zp}
\Delta m  = f (x, y) = C_{0} + C_{1} x + C_{2} y 
\end{equation} where $\Delta m$ is the difference between reference and measured magnitudes for each star on every frame. The magnitude error is used as weighting factor for regression. $C_{0}$ is a good diagnostic for discerning whether unrecognized temporal changes in atmospheric transparency have occurred during the observations. $C_{0}$ should be close to zero when the quality of data is adequate. Figure \ref{fig:jkasfig4} shows the histogram of $C_{0}$ values for all frames along with cut-out images having three different quality cuts (good, intermediate, and bad). Finally, we apply the resulting zero-point surface to the instrumental magnitudes. A simple Boolean index is added to the database to be used as a photometric quality flag of our internal calibrations (see  \texttt{CALDEX} in Appendix). 

% Section 3.2
\subsection{Photometric Recalibration\label{sec:photcal}}
To overcome the absence of observations used for photometric standardization, as mentioned in Introduction, we recalibrate our photometry to the SDSS Data Release 13 (SDSS DR13; \citealt{alb2017}) photometric catalog. The main reason being that there are no suitable photometric catalogs surveyed by other projects (e.g., no overlap with the first public data release of the Dark Energy Survey or poor photometric quality of the third data release of the Palomar Transient Factory). In addition to this, the well-studied SDSS color system is outstanding compared to other photometric systems (e.g., Pan-STARRS $grizy$ filters or even for the Gaia broad bandpass) to aid in the study of point sources identified by our survey. Because only part of the DEEP-South field is covered by the SDSS DR13 photometric catalog, this approach is limited mainly by the declination limit. The recent SkyMapper Southern Survey provides the most comprehensive map of the Southern sky released to the public which contains well-calibrated magnitudes of point sources up to 18 (AB mag) in all $uvgriz$ bands \citep{wol2018}, and thus it can serve as a calibration reference for the full data sets in the future.

\begin{figure}[!ht]
\centering
\includegraphics[width=80mm]{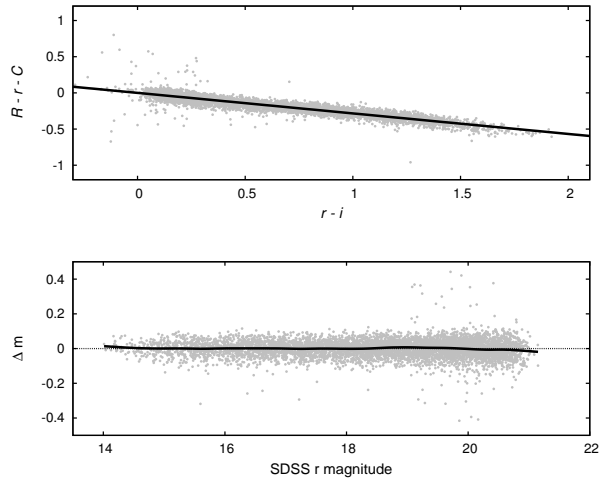}
\caption{Transformation between the DEEP-South $R$ and SDSS $r$ magnitudes as a function of $r - i$. The light points are stars with low photometric error (less than 0.1 magnitudes) in $griz$ bands. Top: the solid line shows the best fitting with Equation \ref{eq:Xtods}; Bottom: corresponding difference between transformed DEEP-South and SDSS magnitudes as a function of the SDSS $r$ magnitude. \label{fig:jkasfig5}}
\end{figure}

From the SDSS DR13 database, we select suitable calibration stars with a series of clean flags (e.g., no edge, non-saturated, no cosmic-ray hits, primary object or no neighborhoods), magnitude errors below 0.05 in $ri$ bands, and PSF magnitudes in the ranges 14$\leq r \leq$20 and 14$\leq i \leq$20. For the DEEP-South data, the mean magnitude error of the sources used becomes less than 0.05 mag in a given magnitude range. The transformation equation between the Johnson $R$ and SDSS $r$ system is simply derived as  
\begin{equation}
\label{eq:Xtods}
R  = r + k (r - i) + C % R = r - 0.2837 (r - i ) + 0.4449 (R=0.94)
\end{equation} where $k$ is the first-order color term and the possible zeropoint shift, $C$, is determined by iteratively rejecting outliers in the residual. The median absolute deviation of the residual of the fit for -0.3 $< r - i <$ 2.1 is 0.029. The color term $k$ equals to -0.2837, which is similar to that found previously ($k$=-0.2936); please refer to the transformation equations derived by Lupton (2005)\footnote{\url{http://www.sdss.org/dr13/algorithms/sdssUBVRITransform/#Lupton2005}}. Figure \ref{fig:jkasfig5} shows the transformation between SDSS $ri$ and DEEP-South $R$ magnitudes as defined by Equation \ref{eq:Xtods}. The scatter of magnitude difference between these two photometric systems is roughly constant with respect to magnitude. We see that the impact of a few photometric outliers is negligible at these color and magnitude levels.

As a final step, we ingest all frame catalogs into the FastBit database (see Appendix for technical details) that provides a set of compressed bitmap indexes to quickly retrieve the list of objects in a given sky region, observational parameters of selected sources or time-series data (e.g., light curves). 

% Section 4
\section{Searching for Temporal and Spatial Variability\label{sec:varsearch}}
\subsection{Light curve production}
Performing a cone search is the easiest way to construct light curves for stationary sources detected at least twice, called \textit{groups}. We first build a master catalog of about 46,000 groups merged by matching positions (with a match radius of 0.5 arcsec) across different epochs and amplifiers, and we also generate a catalog of transient sources detected only once (see Section 4.3). Our smaller matching radius can reduce the number of spurious matches, but it causes us to lose a few real matches for faint sources. We find that about 80 new groups (mostly $R$ $>$ 20) can be associated by positional matching if we double the cutoff radius, thus the group sample considered here may not severely limit our variability results in the next section. Within each group, we determine mean positions by combining all astrometric measurements after excluding flagged data points. Now we only require the mean sky position and a selected search radius to produce a light curve of each group. We choose a loose cut for light curve production (i.e., matching radius of 0.8 arcsec). The following example command executes a query on the database and returns associated measurements:
\begin{verbatim}
ibis -d DeepSouth-DB
     -q "SELECT MJD, MAG_MAP, MAGERR_MAP
        WHERE (SQRT(POWER(XWIN_RA-RA_group, 2) 
             + POWER(YWIN_DEC-DEC_group, 2))
        BETWEEN 0 AND SEARCH_RADIUS)
        AND FLAGS < 4 
        ORDER BY MJD".
\end{verbatim} In this paper, we limit the sample to $\sim$13,000 stars which have reliable color information in the SDSS DR13 database. Note that there is overlapping regions with the Pan-STARRS1 data, but it is similar in depth to the SDSS in $r$-band for point sources. %Metcalfe et al. MNRAS 435, 1825 (2013) has a detailed comparison of the PS1 Small Area Survey 2 data (a test 3pi region taken under optimal conditions) with SDSS.

\begin{figure}[!t]
\centering
\includegraphics[width=85mm]{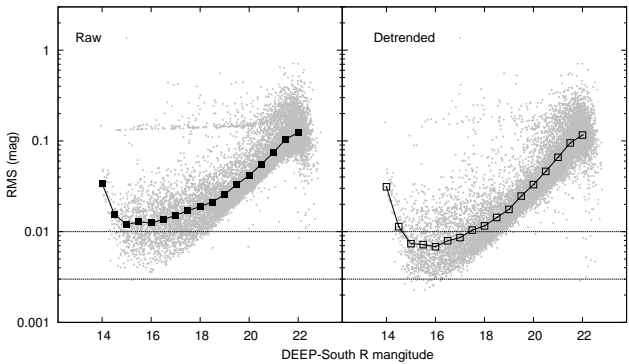}
\caption{Robust RMS of light curves before (left) and after (right) the photometric detrending as a function of magnitude. The solid lines are  smoothed averages using a bin width of 0.5 magnitude. The horizontal lines indicate a photometric precision of 1\% and 0.3\%, respectively.\label{fig:jkasfig6}}
\end{figure}

In order to remove systematic noise caused by atmospheric variations or instrumental effects, we apply the photometric detrending algorithm (see \citealt{kim2009} for details)\footnote{\url{https://github.com/dwkim78/pdtrend}} to the light curves. This algorithm finds position- and time-dependent systematics using a clustering technique, and then corrects the highly affected light curves by removing the systematics from individual light curves. Figure \ref{fig:jkasfig6} shows the overall dispersion of light curves before and after the detrending procedure without any outlier clipping. Here, we use the root-mean-square (RMS) amplitude of light curves as a proxy of our photometric precision. The RMS values of DEEP-South light curves decrease to 0.3\% precision level at the bright end ($R$ $\le$ 16). Applying this technique to the raw light curves gives good results in recovering the true brightness variations over the full magnitude range from 14 to 22.5.

\begin{figure}[!t]
\centering
\includegraphics[width=80mm]{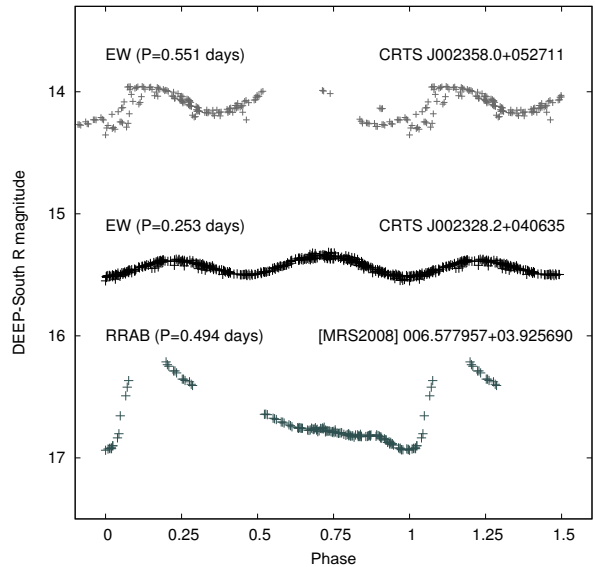}
\caption{Final light curves of three known variable stars folded by their period. Both variable types and periods are noted above each phased light curve.\label{fig:jkasfig7}}
\end{figure}

\begin{figure*}[!t]
\centering
\includegraphics[width=80mm]{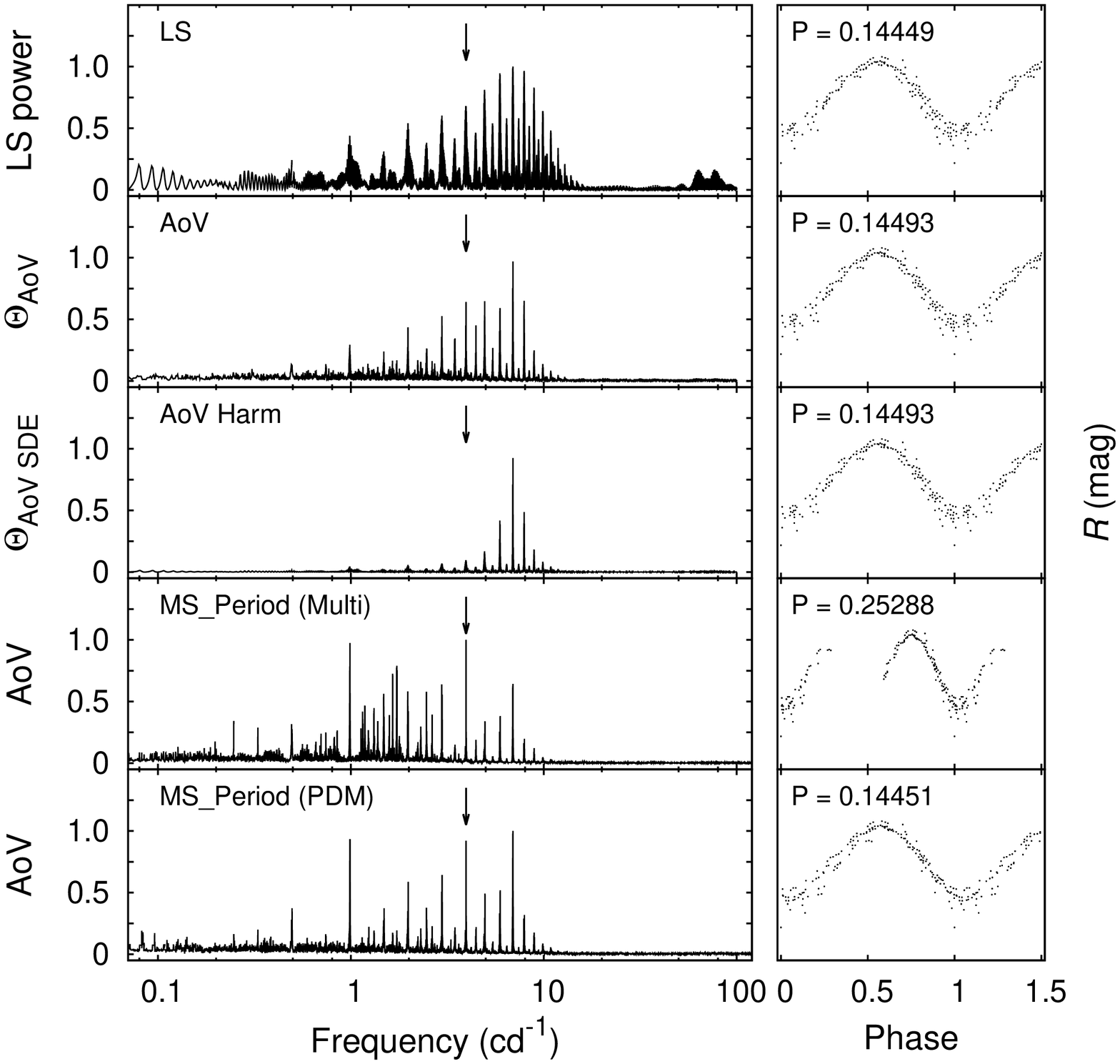}
\includegraphics[width=80mm]{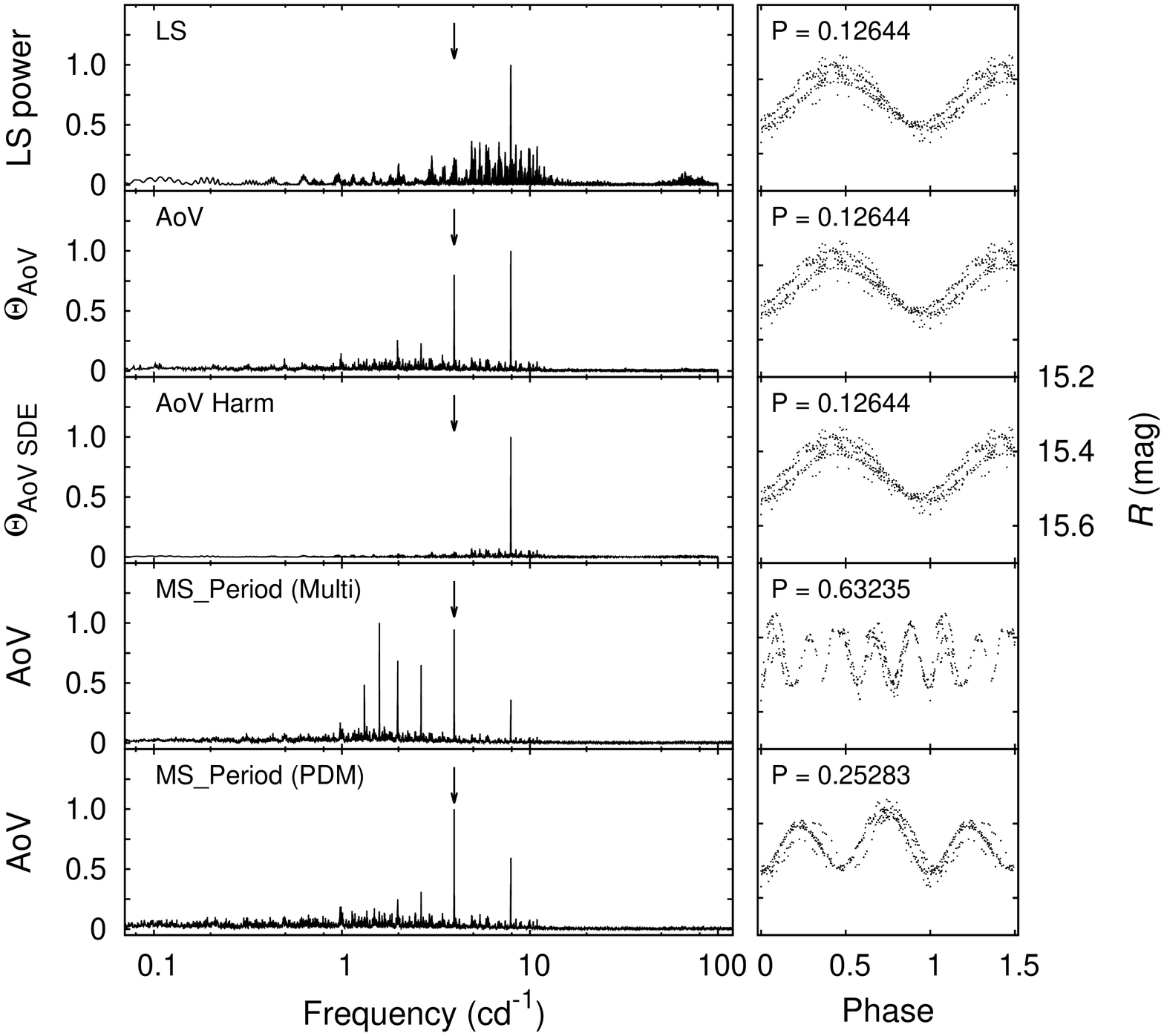}
\caption{Period finding results by various periodogram tools for the CTIO sample only (left) and for the combined sample from three telescopes (right). Each periodogram power is normalized by the peak amplitude for comparison purpose. The arrows indicate a true period of 0.25288 days for one known EW variable star shown in Figure \ref{fig:jkasfig7}.\label{fig:jkasfig8}}
\end{figure*}

Figure \ref{fig:jkasfig7} shows the detrended, final light curves of known variable stars from the AAVSO International Variable Star Index (VSX: \citealt{wat2017}). We identify three out of five known variable stars in the N02007-OC field; two of them are W Ursae Majoris-type eclipsing variables (EW) with periods shorter than 0.5 days (CRTS J002358.0+052711 and CRTS J002328.2+040635). The other one is a typical ab-type RR Lyrae star, [MRS2008] 006.577957+03.925690, with heliocentric distance $\sim$15.84 kpc. The photometric quality of folded light curves is good enough to conduct variability analysis. One star shown in Figure \ref{fig:jkasfig7}, CRTS J002358.0+052711, is close to the saturation limit ($\approx$14--15 mag depending on seeing conditions)  in the N02007-OC field causing the relatively large scatter of the light curve over all phase bins. Two of the unmatched variables (ASAS J002137+0513.8 and CD Psc) are bright ($<$12 mag) and saturated in our observations.

\subsection{Periodogram analysis for variability search}
One explicit advantage of longitudinal network observations with the KMTNet facility is to alleviate aliasing signals due to the common occurrence of data gaps in unevenly spaced time-series data. Figure \ref{fig:jkasfig8} compares the results of period analysis by various periodogram tools (e.g., VARTOOLS\footnote{\url{http://www.astro.princeton.edu/~jhartman/vartools.html#Home}}: \citealt{har16} and MS\textunderscore Period\footnote{\url{https://astromsshin.github.io/science/code/MultiStep_Period/index.html}}: \citealt{shin2004}) for the CTIO sample only and for the combined data from three telescopes. The spectral peaks in the periodogram are severely affected by daily aliasing signals for the CTIO sample, but combining the time series data from the three sites greatly reduces the false peaks due to aliasing.

\begin{figure}[!ht]
\centering
\includegraphics[width=80mm]{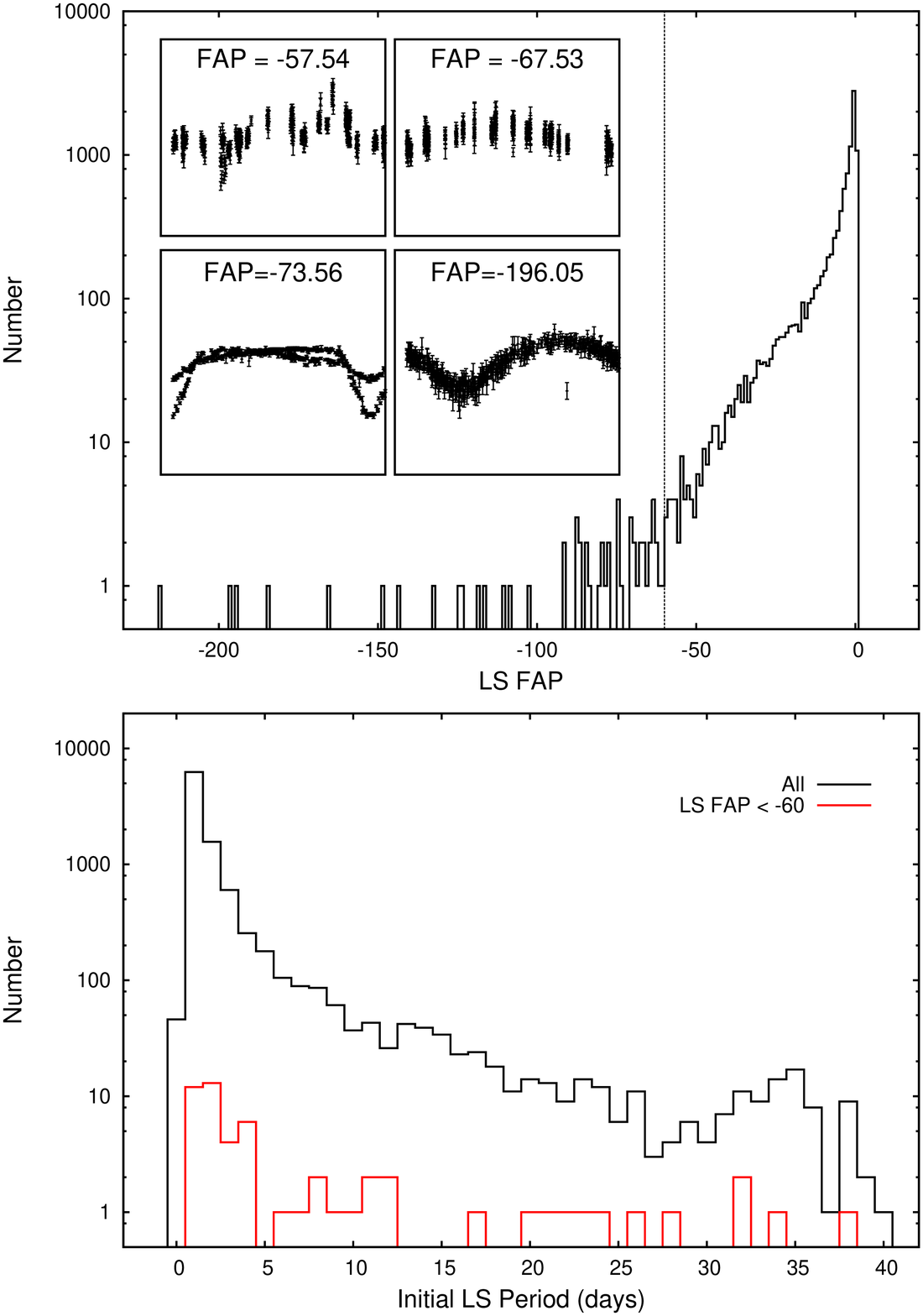}
\caption{Histograms of the FAP values (top) and initial periods (bottom) from the LS algorithm after filtering alias signals. Our conservative selection criterion is LS FAP $<$ -60 indicated by dashed line. Example phased light curves (magnitude versus phase) are also shown for guiding purpose. \label{fig:jkasfig9}}
\end{figure}

We mainly use two different algorithms implemented in the VARTOOLS light curve analysis program to search for periodic variables with periods less than 40 days at an initial frequency resolution of 0.1/$T$ ($T$ is the time-span of the light curve): a Generalized Lomb-Scargle (LS) algorithm and an Analysis of Variance (AoV) algorithm using either phase binning or multi-harmonic model fitting (see \citealt{har16} and references therein). To reduce erroneous results caused by outliers, we applied a typical sigma clipping (4-sigma) to each light curve before searching for semi-sinusoidal signals. Each algorithm gives a diagnostic parameter to test the significance of candidate periods. We choose different conservative selection cuts based on a false alarm probability (FAP) by visual inspection of example light curves at a range of values: e.g., LS FAP $<$ -60, AoV FAP $<$ -100, and AoVharm FAP $<$ -140, respectively. Moreover, we remove the most obvious spurious detections having a value of peak frequency very close to 1 or 2 c/d (mostly around $\sim$0.99 and  $\sim$2.02 days) due to the daily gaps even in the combined sample (see Table \ref{tab:jkastable1} for observation dates). Figure \ref{fig:jkasfig9} shows the histograms of formal FAP values and initial periods from the LS period-search algorithm as an example after removing aliases. With these criteria, we check all the phase-folded light curves by eye and refined the initial estimate of the periods using fine grid searches ($<$0.01/$T$) around the highest peaks. Table \ref{tab:jkastable3} lists 21 new periodic variable stars in the period range of 0.1--31 days. From most stars, the difference, $\Delta P$, in the periods computed by using the LS and AoV algorithms is less than 0.5\%. Table \ref{tab:jkastable2} summarizes the steps we used to identify a sample of periodic variable stars from our database for the N02007-OC field.

\begin{table}[t!]
\caption{Periodic Variable Selection Criteria \label{tab:jkastable2}}
\centering
\begin{tabular}{cc}
\toprule
Selection criterion & Number of objects \\
\midrule
All 2.5-$\sigma$ sources & 15,629,598$^{\rm a}$  \\
All possible groups ($<$0.5 arcsec)  & 46,054 \\
SDSS stars with $gri$ colors & 13,261 \\
Variable candidates by FAP cuts & 309 \\
Rejection of aliased signals & 57 \\
Final periodic variables   & 24 \\ %\addlinespace
\bottomrule
\end{tabular}
\tabnote{ $^{\rm a}$ Number of individual photometric measurements.}
\end{table}

\begin{figure}[!ht]
\centering
\includegraphics[width=75mm]{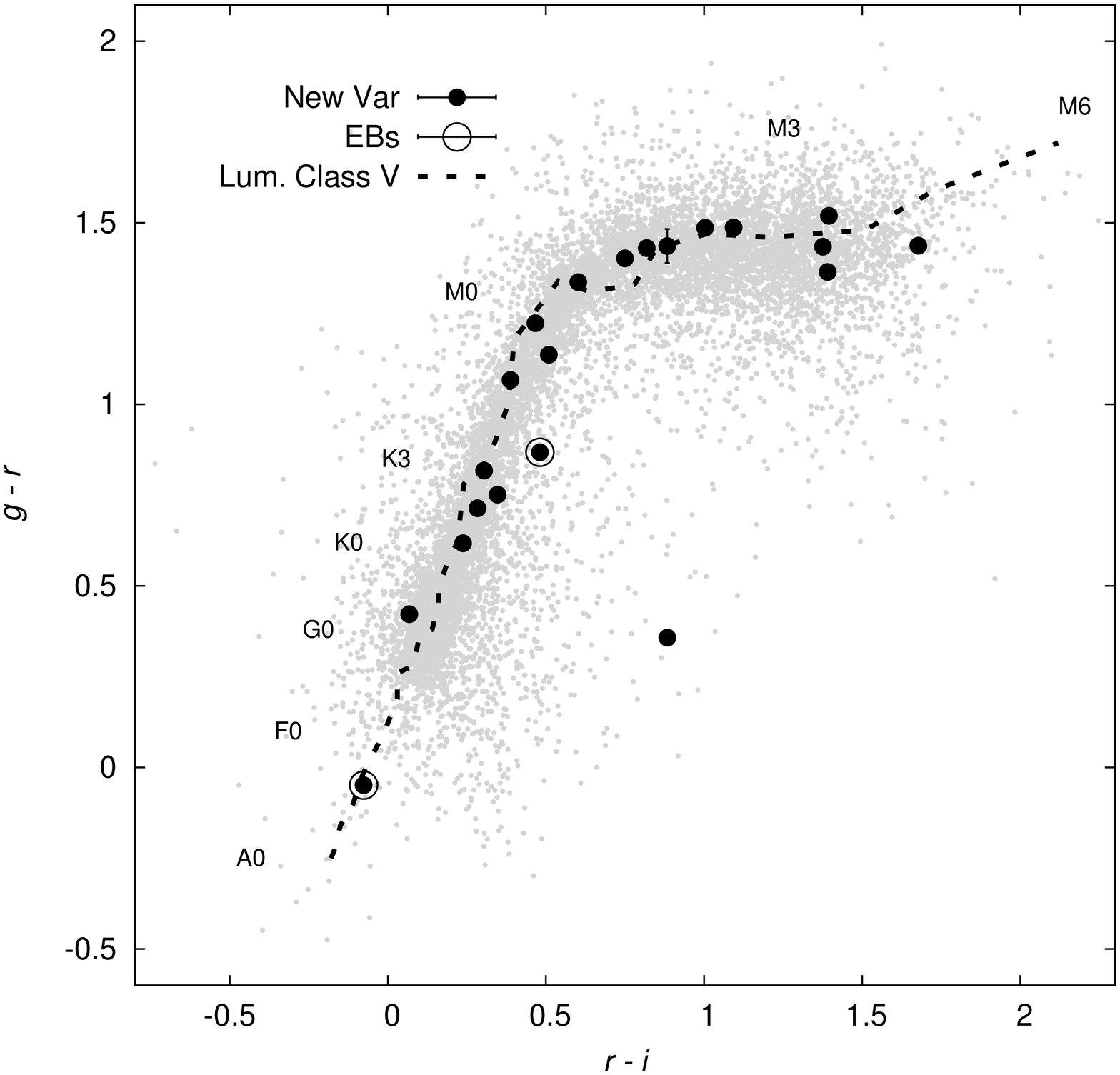}
\includegraphics[width=75mm]{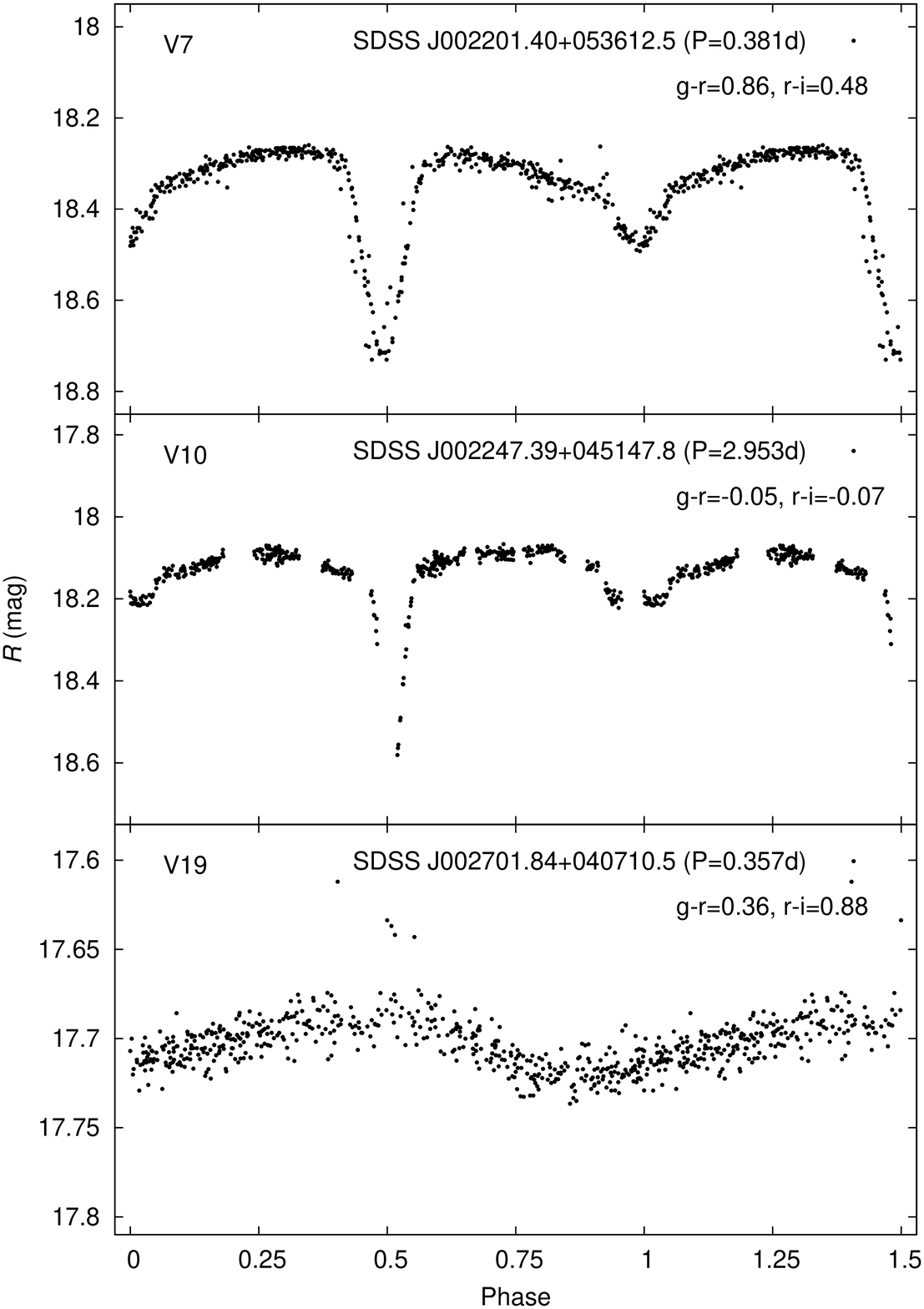}
\caption{Location of 21 new periodic variable stars (black dots) and all other stars (gray dots) in color-color diagram. The dashed line indicates the main-sequence stellar locus with solar metallicity \citep{cov2007}. The locations of two eclipsing binaries and one WDMD candidate are indicated by circles and square with overplotted black dots, respectively. The bottom panels show their phase-folded light curves.
\label{fig:jkasfig10}}
\end{figure}

\begin{table*}
\caption{21 New Periodic Variable Stars in the N02007-OC field\label{tab:jkastable3}}
\begin{tabular}{ccccccccccc}
\toprule
VarID & StarID & $R$ & $P$ & $A_{R}^{\rm a}$ & $u$ & $g$ & $r$ & $i$ & $z$ &  Note\\
& (J2000 coordinates)  & (mag) & (days) & (mag) & (mag) & (mag) & (mag) & (mag) & (mag) & \\
\midrule
V1 &  SDSS J002013.16+052242.2 & 16.24 & 2.1038 & 0.06 & 19.66 & 17.35 & 16.21 & 15.71 & 15.35 &  \\
V2 &  SDSS J002031.74+055112.6 & 16.33 & 1.1568 & 0.04 & 20.26 & 17.72 & 16.38 & 15.78 & 15.43 &  \\
V3 &  SDSS J002051.80+042641.3 & 17.50 & 0.3076 & 0.13 & 19.59 & 18.10 & 17.48 & 17.25 & 17.15 &  ELV \\
V4 &  SDSS J002111.05+054856.4 & 18.60 & 1.5046 & 0.04 & 22.12 & 20.04 & 18.61 & 17.23 & 16.46 &   \\
V5 &  SDSS J002125.21+040357.0 & 16.79 & 31.5771 & 0.03 & 21.12 & 18.20 & 16.80 & 16.05 & 15.62 &  \\
V6 &  SDSS J002141.37+051626.1 & 16.76 & 22.4097 & 0.01 & 20.83 & 18.14 & 16.71 & 15.89 & 15.47 &   \\
V7 &  SDSS J002201.40+053612.5 & 18.35 & 0.3808 & 0.44 & 21.18 & 19.28 & 18.41 & 17.93 & 17.60 &  EB \\% 120129 
V8 &  SDSS J002215.91+055338.9 & 20.19 & 0.2489 & 0.32 & 21.56 & 20.68 & 20.26 & 20.19 & 20.09 &  OC\\
V9 &  SDSS J002216.31+040439.5 & 19.59 & 23.4399 & 0.12 & 23.36 & 21.13 & 19.69 & 18.01 & 17.05 &   \\
V10 &  SDSS J002247.39+045147.8 & 18.11 & 2.9532 & $>$0.49 & 19.19 & 18.04 & 18.09 & 18.17 & 18.27 &  EB  \\% 270161
V11 &  SDSS J002249.03+055356.6 & 18.33 & 16.0571 & 0.04 & 22.39 & 19.82 & 18.34 & 17.33 & 16.79 &   \\
V12 &  SDSS J002314.27+042543.6 & 19.00 & 0.2673 & 0.16 & 23.05 & 20.33 & 18.97 & 17.58 & 16.75 &   \\
V13 &  SDSS J002351.64+054553.7 & 15.89 & 21.7415 & 0.02 & 19.84 & 17.10 & 15.88 & 15.41 & 15.12 &  \\
V14 &  SDSS J002425.52+044130.3 & 17.95 & 1.6484 & 0.10 & 20.22 & 18.64 & 17.89 & 17.54 & 17.38 &  \\
V15 &  SDSS J002435.70+051914.1 & 16.64 & 7.5176 & 0.03 & 19.04 & 17.30 & 16.59 & 16.30 & 16.13 &  \\
V16 & SDSS J002516.12+053728.4 & 15.09 & 11.9412 & 0.02 & 18.82 & 16.14 & 15.07 & 14.68 & 14.46 &  \\
V17 &  SDSS J002549.95+041608.6 & 20.60 & 0.1234 & 0.10 & 23.56 & 21.98 & 20.54 & 19.66 & 19.13 &  \\
V18 &  SDSS J002610.07+043147.2 & 18.15 & 2.8856 & 0.02 & 22.81 & 19.64 & 18.12 & 16.73 & 15.95 &    \\
V19 &  SDSS J002701.84+040710.5 & 17.69 & 0.3570 & 0.03 & 18.46 & 18.02 & 17.66 & 16.78 & 16.15 &  WDMD \\% 190084
V20 &  SDSS J002704.93+040519.4 & 15.32 & 20.4821 & 0.02 & 18.45 & 16.20 & 15.38 & 15.08 & 14.89 &   \\
V21 &  SDSS J002733.09+043701.9 & 17.70 & 3.1778 & 0.04 & 21.41 & 19.15 & 17.66 & 16.57 & 15.99 &  \\
%J002328.26+040635.1 & 15.43 & 0.2529 & 0.05 &  EW \\ % 260010 EW (Known) %https://academic.oup.com/view-large/figure/64787360/stx567fig3.jpg/When%20flux%20standards%20go%20wild%3A%20white%20dwarfs%20in%20the%20age%20of%20Kepler
\bottomrule
\end{tabular}
\tabnote{
$^{\rm a}$  Peak-to-through $R$-band variability amplitude of sinusoidal function fitted to phase-folded light curve.
}
\end{table*}

\begin{figure*}[!h]
\includegraphics[width=85mm]{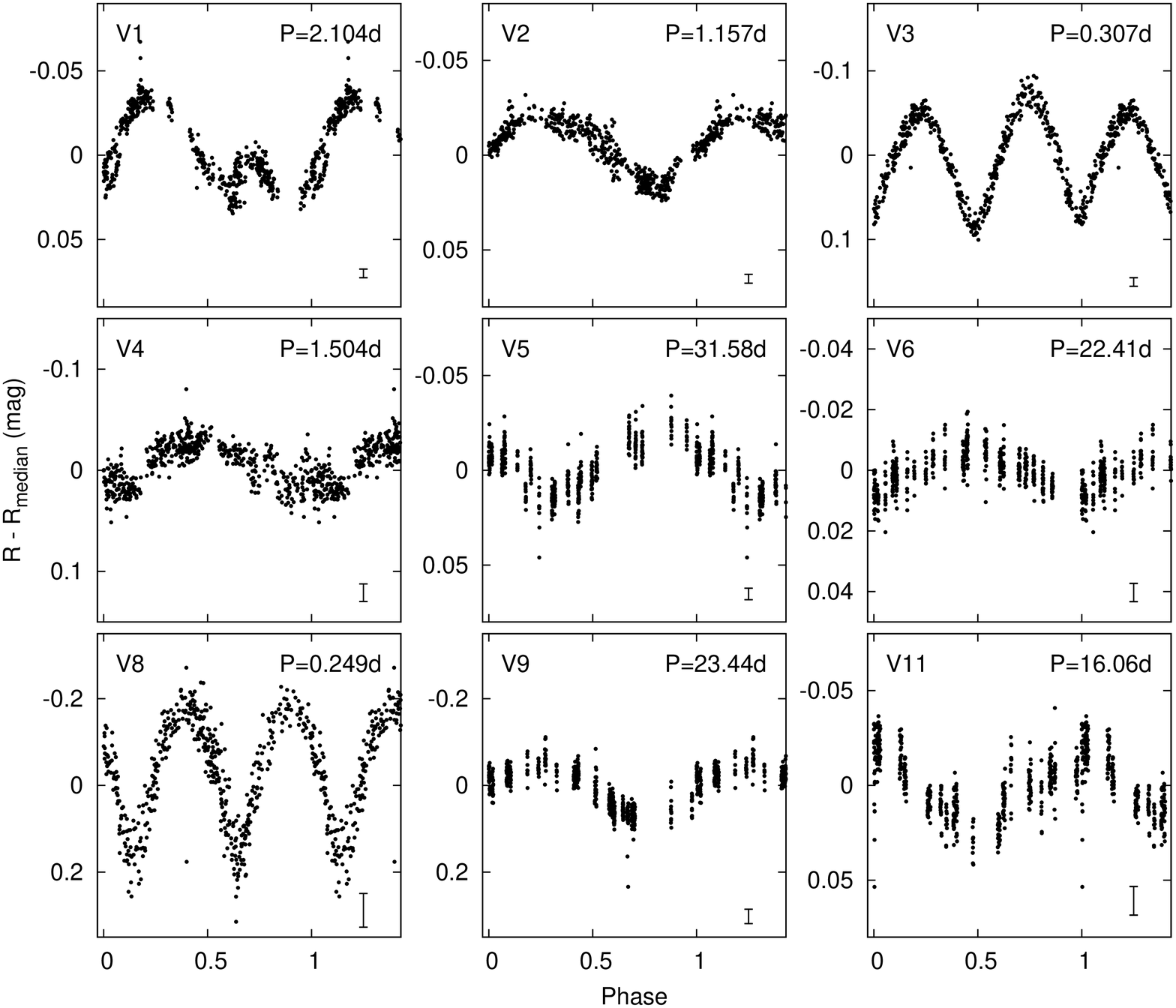}
\includegraphics[width=85mm]{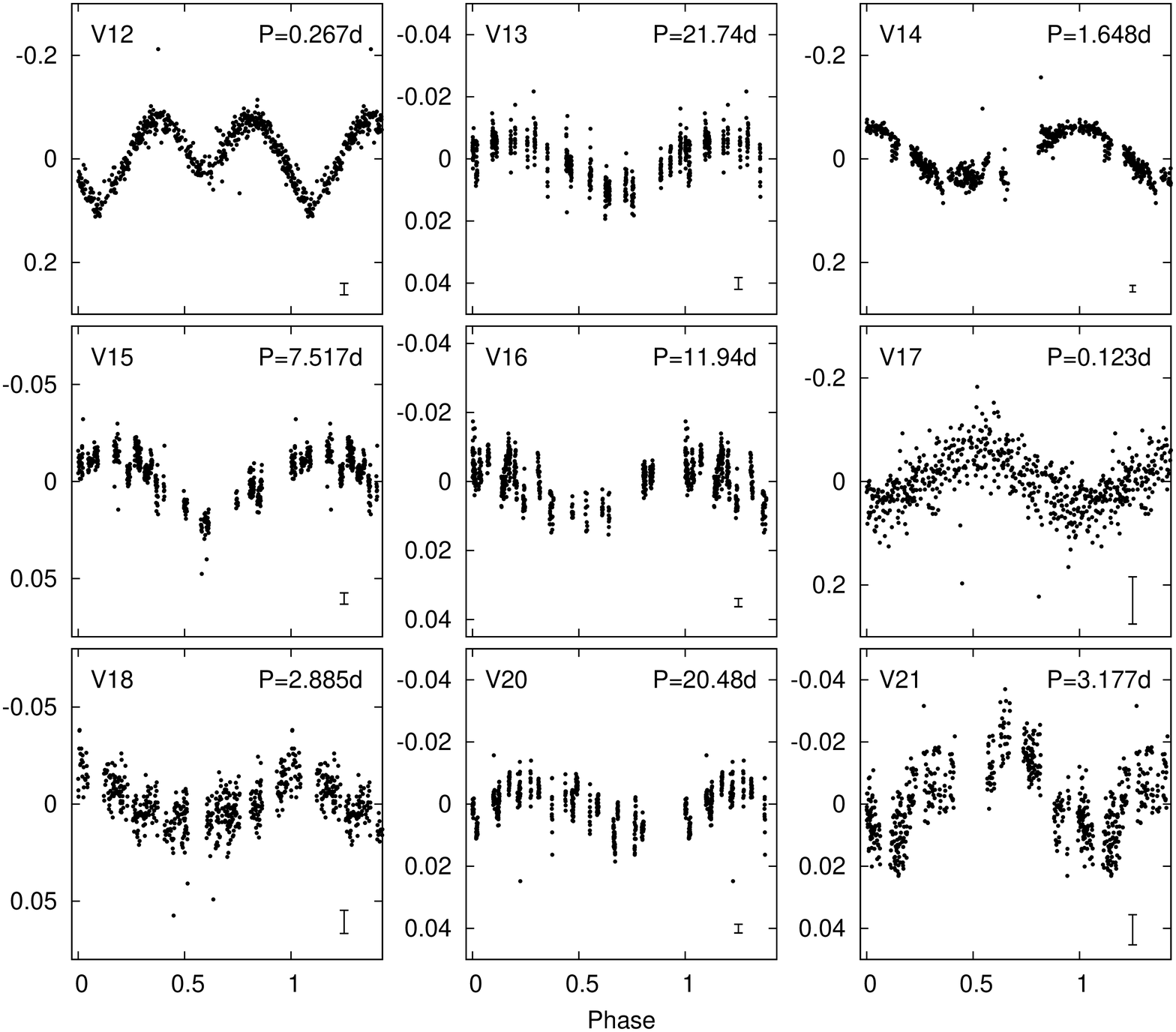}
\caption{Phase-folded light curves of other 18 periodic variable stars. The small error bars indicate the interquartile range of the measurement errors for each light curve. 
 \label{fig:jkasfig11}}
\end{figure*}

Figure \ref{fig:jkasfig10} shows the locations of newly discovered variable stars in a color-color diagram, as well as all other stars with magnitude error less than 0.2 in $gri$ bands. The effect of SDSS color errors induced by variability is negligible for our variable stars (error bars are smaller than the symbols). Most of them are close to the locus of stellar main sequence described by the SDSS $gri$ colors (\citealt{cov2007}, see dashed line in the same figure). We have only limited information to characterize their variability nature as their spectra are currently not available yet. However, the lightcurve properties and source colors suggest that these are mainly due to spot-induced rotational modulation. One color outlier that lies far outside the locus is a candidate white dwarf/M dwarf pair (WDMD: SDSS J002701.84+040710.5; V19), which is a likely short-period binary showing reflection from a close companion at the orbital period.  The observed color is also consistent with those expected from genuine WDMD binaries (e.g., see Figure 2 of \citealt{reb13}).  As shown in Figure \ref{fig:jkasfig10}, two of them show eclipsing binary (EB) signatures in their phase-folded light curves. The epochs of the primary and secondary eclipses are clearly seen over more than one complete cycle. SDSS J002201.40+053612.5 (V7) is identified as semi-detached EB with a 0.3808 d orbital period. SDSS J002247.39+045147.8 (V10) is an Algol-type EB with an orbital period of 2.953 days, having the detached component. We also show phase-folded light curves of all other discovered variables in Figure \ref{fig:jkasfig11}. SDSS J002215.91+055338.9 (V8) is a overcontact (OC) binary which is recognized by a continuous variation between eclipses such as W UMa-type systems. Lastly, SDSS J002051.80+042641.3 (V3) may be an ellipsoidal variable (ELV) showing double maxima and minima per orbital period due to tidal distortion. 

\begin{figure*}[!ht]
\includegraphics[width=80mm]{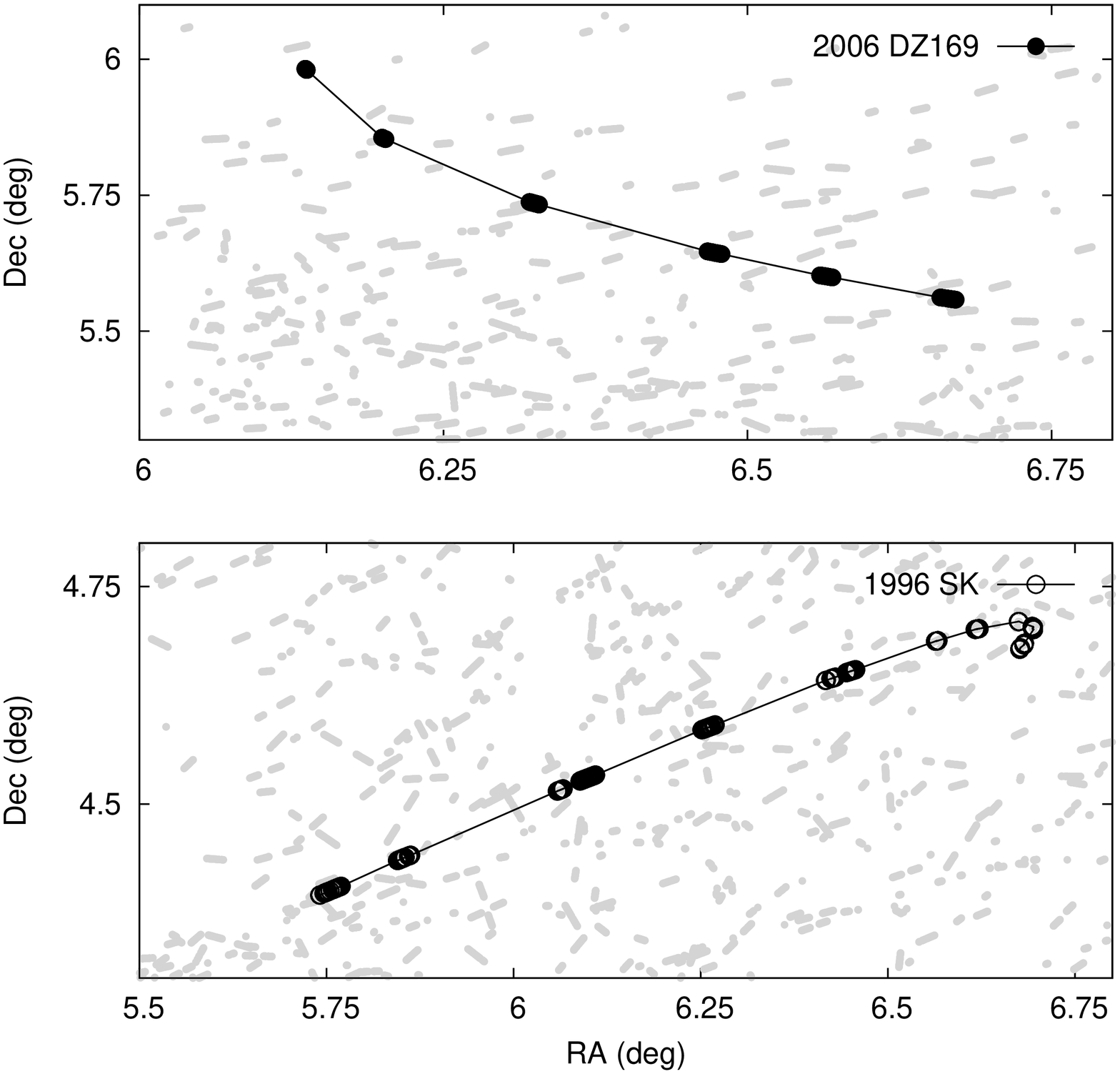}
\includegraphics[width=80mm]{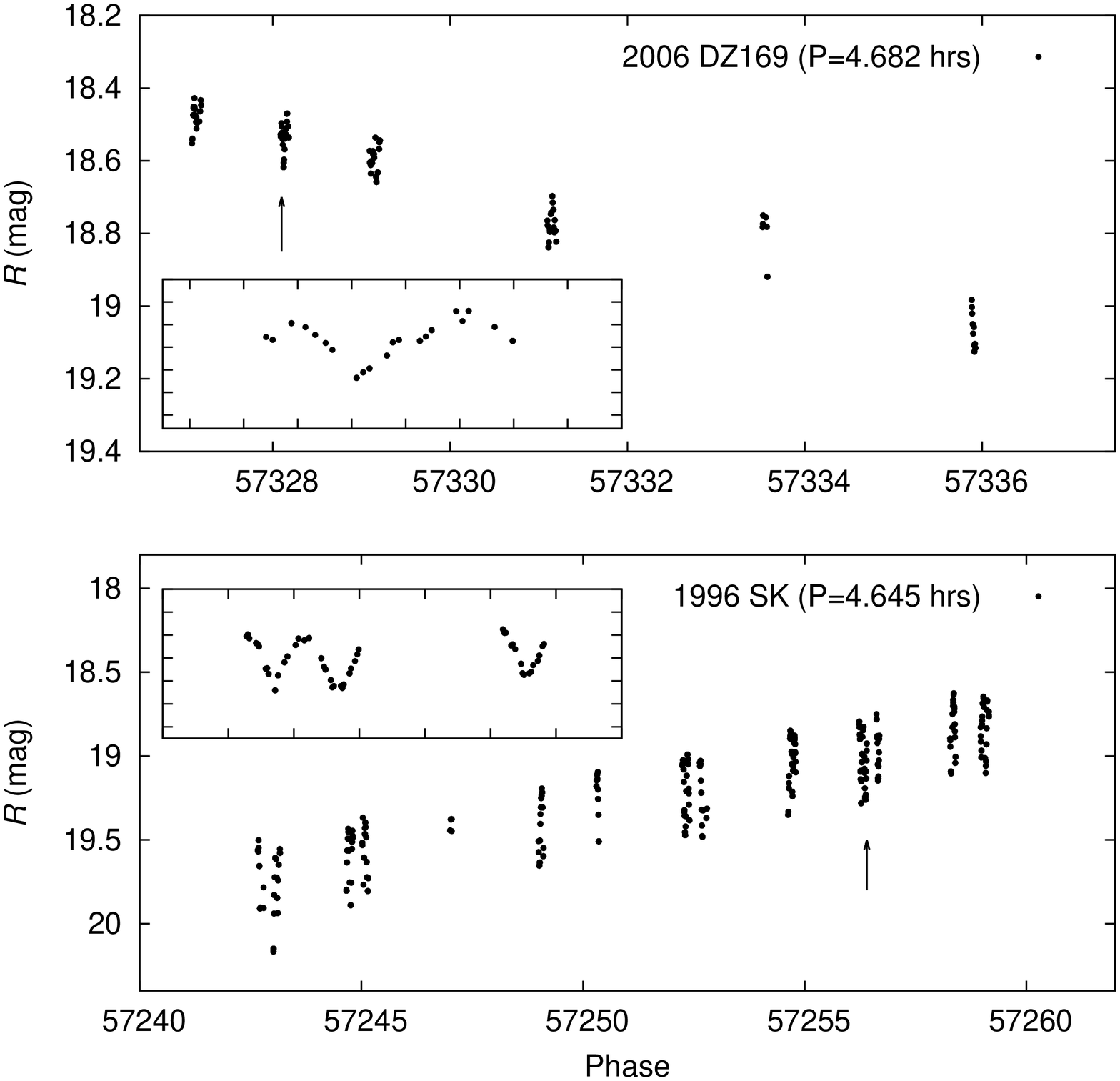}
\caption{Projected orbital paths (left) and light curves (right) of 2006 DZ169 and 1996 SK. Left: The background gray dots are single-epoch sources in the transient catalog, showing only known moving objects listed in the VO SkyBoT database. Right: The arrows indicate a subset of light curves zoomed in the inserted plots which show rapid (less than a few hours) brightness variations due to rotation. \label{fig:jkasfig12}}
\end{figure*}

% http://earn.dlr.de/nea/155334.htm
% http://earn.dlr.de/nea/297274.htm
\subsection{Recovery of moving objects}
Another application is to retrieve trajectories of targeted asteroids and their light curves in the catalog of transient sources, i.e., those detected only once. It is not difficult to recover orbital paths of known asteroids on every frame by comparing with an ephemeris from the Minor Planet Center (MPC)\footnote{\url{https://www.minorplanetcenter.net/}}, the official organization that is responsible for the identification, designation and orbit computation for all types of moving objects (minor planets, comets and outer irregular natural satellites of the major planets). The names of the two objects we observed in this field are 2006 DZ169 and 1996 SK. The former asteroid was considered as a potential target for space missions due to its low rendezvous velocity (e.g., \citealt{mue11}), while the latter is a potentially hazardous NEA (e.g., \citealt{lin14}). Their physical and dynamical properties are well summarized in the NEAs database, updated continuously by European Asteroid Research Node. There is no ambiguity in their measured rotation periods with a reliability code of 3 (i.e., secure result with full lightcurve coverage) based on the definition of \citet{lag1989}. Therefore, our observations allow us to confirm the results of variability analysis reported in previous works\footnote{\url{https://ssd.jpl.nasa.gov/sbdb_query.cgi}}.

% number of single-epoch sources: 113,241
We compute celestial coordinates of these NEAs at a given epoch based on the orbital data provided by the MPC database, and then we cross-identify single-epoch sources in the transient catalog with those pre-computed positions. In this way, we recover 98\% of the spatial locations of the NEAs after excluding flagged data. The coordinate difference between the MPC and our astrometric solution is less than $\pm$1--2 arcsec in both RA and Dec direction.  Figure \ref{fig:jkasfig12} shows the projected path of each orbit (solid lines) over the observing span, as well as the single-epoch sources that are detected in this region of the sky. In order to highlight the possibility of searching for untargeted moving objects with moving speed similar to targeted ones, we only include all known moving objects within a given FoV (gray linked steaks or points)  identified by the Virtual Observatory SkyBoT tool \citep{berth2006}.

Lastly, we present light curves of the two targeted NEAs which are representative in quality for other asteroids in the transient catalog because our experimental data were obtained in conditions optimal to measure their periods (see Figure \ref{fig:jkasfig12}). These light curves are a superposition of two components with different timescales; a long-term change in brightness with increasing or decreasing solar phase angle and a rapid periodic modulations in brightness due to rotation of the asteroid (see zoomed-in views of both light curves). After removing the long-term light variation and outliers, we measure rotation periods and full amplitudes for the NEAs using the periodogram tools. Their rotation periods are found to be similar to those in the NEAs database. The rotation period of 1996 SK is about 4.644 hours ($ P_{new}$) with a highly reliable result with full light curve coverage (c.f.,$ P_{known}$=4.645 hrs), while that of 2006 DZ169 is about $ P_{known}$=4.682 hours that is not clearly seen in the data. 1996 SK has a large amplitude of $\sim$0.5 mag in $R$-band, while 2006 DZ169 exhibits a relatively low amplitude variation ($\sim$0.12 mag).

%-1996 SK: 4.645 hrs (lightcurve amplitude: 0.39-0.45 mag) vs. 4.644384 hrs + our $A_{R}$ is $\sim$0.5 mag
%-2006 DZ169: 4.682 hrs (0.21 mag) vs. similar period but not clear in our data + $A_{R}$ is $\sim$ 0.12 mag

%The 4th Release of the Sloan Digital Sky Survey Moving Object Catalog \url{http://faculty.washington.edu/ivezic/sdssmoc/sdssmoc.html}; mean (r -i) color of moving objects is 0.3137, so we have to add 0.0889; R +0.2837 (r - i ) - 0.4449  = r

% Section 5
\section{Summary\label{sec:summary}}
We have described in detail the main algorithms of our new reduction pipeline to explore the temporal and spatial variability with DEEP-South observations. Our multi-aperture photometry technique produces a homogeneous set of photometric measurements for all point sources in non-crowded fields observed by a distributed network of three different telescopes. This is important as it will allow us to study the variability properties of targeted or untargeted objects in a large database of photometric time series without extra computational efforts. Taking into account spatial dependence of PSF variations, zeropoint variations, and systematic effects, we find the RMS scatter of the light curves for point sources to be reached down to 0.3\% level at the bright end ($R$ $\le$ 16) and $\sim$10\% level at the faint end ($R$ $\sim$ 22) in the longest observing records of the DEEP-South year-one data. We also emphasize the use of the indexed database tools, such as FastBit, in both minimizing required storage space and fast query performance to produce large sets of light curves. In the future, we will apply this updated pipeline to process entire imaging data taken by DEEP-South survey to further explore stellar variability.

As applications of the photometric database, we first presented light curves of known variable stars to illustrate the overall quality of the photometric calibration. We find 21 new periodic variable stars with period between 0.1 and 31 days, including four EBs and one WDMD candidate which are evident by either (i) the shape of phase-folded light curve and/or (ii) colors to figure out variability classes. Since we limit the samples to sources only with the SDSS colors, we could miss either non-periodic variables or variable candidates without SDSS colors. To fully investigate properties of all remaining sources, we will attempt to use an infinite gaussian mixture model for detecting variable objects and suppressing false positives efficiently (e.g., \citealt{shin2009,shin2012}).

Additionally, we show the potential of database applications to retrieve the projected orbital paths and light curves of two targeted NEAs (2006 DZ169 and 1996 SK) in the experimental data. We expect to see more known asteroids with moving speed similar to targeted ones, and we will present results of exploring this possibility in the second paper of the series. All light curves of variable objects in this experiment can be accessed through the web site: \url{http://stardb.yonsei.ac.kr/}.

% Acknowledgments
\acknowledgments
We thank two anonymous referees for their constructive comments that improved this paper. This research was supported by the Korea Astronomy and Space Science Institute (KASI) under the R\&D program (Project No.2015-1-320-18) supervised by the Ministry of Science, ICT and Future Planning. S.-W. C. acknowledges the support from KASI -- Yonsei research collaboration program for the frontiers of astronomy and space science (2016-1-843-00). Parts of this research also were conducted by the Australian Research Council Centre of Excellence for All-sky Astrophysics (CAASTRO), through project number CE110001020. This research has made use of the KMTNet system operated by the KASI and the data were obtained at three host sites of CTIO in Chile, SAAO in South Africa, and SSO in Australia. 

\appendix

\subsection{Database System\label{sec:fastbit}}
The key technology of the FastBit is a bitmap-based indexing scheme that stores a list of row identifiers for each value of attribute column as sequences of bits (i.e., 0 or 1). It also reduces the size of index file significantly. Due to query-intensive nature of our experiments, this kind of indexing helps improve query performance. \citet{liu2014} reported good performance in index creation, query and storage space of this algorithm by comparing with a typical relational database (e.g., MySQL or PostgreSQL). The reader is referred to \citet{liu2014} for details of quantitative comparison between both database systems using high-resolution solar observations.

\begin{figure}[!t]
\centering
\includegraphics[width=78mm]{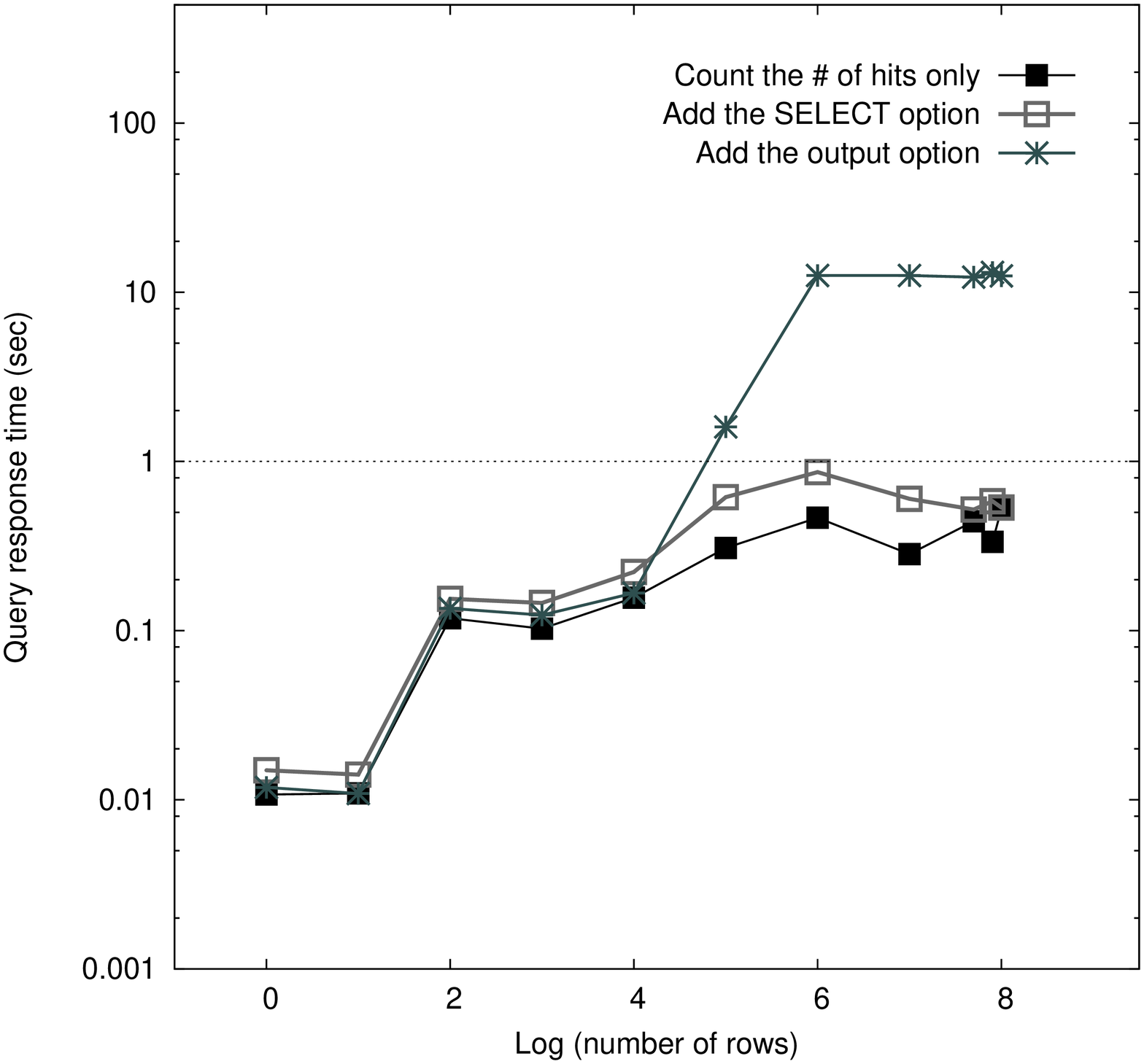}
\includegraphics[width=78mm]{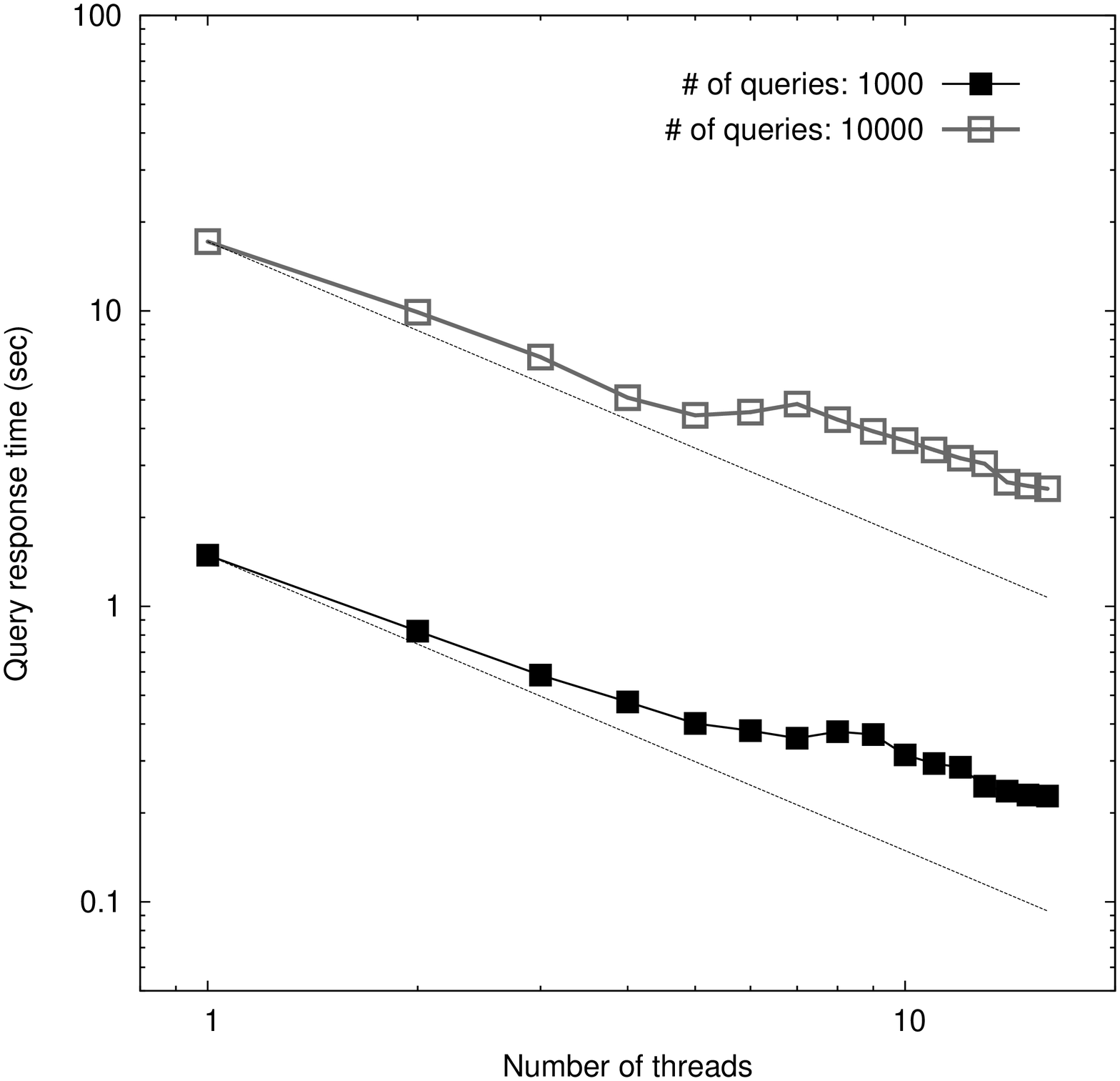}
\caption{Assessment of query execution performance using the FastBit database system for both single-thread (top) and multi-thread configurations (bottom). The dashed lines indicate the slope of perfect scaling with respect to single-node performance for reference. \label{fig:jkasfig13}} 
\end{figure}

We check query response time for two cases: (i) a single one-thread application for different query options and (ii) multi-thread application for multiple query processing. In the former case, we compare performance of counting the number of hits returned without any options to that with only SELECT clause and that with an output option, respectively. For sizes up to 100 million rows, typical queries take less than one second on an Intel Xeon Processor with 2.66GHz clock speed (see top panel of Figure \ref{fig:jkasfig4}). FastBit database operation is limited by I/O performance that is dominated by data-accesses to disk than computations performed by the CPUs. In the multi-thread case, we made either 1000 or 10000 queries to be contained in two files where each SELECT statement is randomly defined. We can see improvement in speed by creating multiple query processes and executing these multiple queries in parallel, as shown in the bottom of Figure \ref{fig:jkasfig13}. We briefly present an example application in an astronomical context; therefore, we request the reader refer to \citet{chou2011} for more details about FastBit-based parallel query processing.

We use command-line tools provided by FastBit for converting data format and building indexes. We hope this will make it easier for users of their own data sets to find relevant content. The following example command converts frame catalogs to the metadata tables in a raw binary form:
\begin{verbatim}
ardea -d DeepSouth-DB (directory-to-write-data)
      -m "LOCAL_ID:uint, XWIN_IMAGE:double, 
          YWIN_IMAGE:double, XWIN_RA:double, 
          YWIN_DEC:double, MJD:double, 
          MAG_MAP:double, MAGERR_MAP:double,
          FWHM_IMAGE:float, ELONGATION:float, 
          CLASS_STAR:float, FLAGS:short, 
          CALDEX:int, AMPS:int"
       -t frame-catalog (text-file-to-read)
       -v 5 (verbose_level)
       -b `'(break/delimiters-in-text-data).
\end{verbatim} The below example command builds new indexes with basic bitmap options for binning, encoding, and compressing processes:  
\begin{verbatim}
ibis -d DeepSouth-DB (target database)
     -b "<binning nbins=B/><encoding equality/>" 
     -z (append-existing-indexes)
     -v 5 (verbose_level).
\end{verbatim} We choose the binning option to reduce the number of bitmaps for attributes with (very) high cardinalities. The strategy of binning was discussed in detail in section 2.4 of \citet{wu2009}, where it is shown that binning can improve the query response time based on order-preserving bin-based clustering. The maximum size of the index is primarily determined by three parameters: the number of rows $N$, the number of bins $B$, and the bitmap encoding. Under the equality encoding condition, our test database contains 15,629,598 rows each with 17 attribute columns, and its indexed size is about 2.4 gigabytes for small $B$ ($<$ 100).

Unfortunately, the use of FastBit scheme has intrinsic limitations as it is just a stand-alone data processing tool. It is not a database management system, so most SQL commands are not supported. It imposes a limit on the number of rows that can be stored in indexed tables (no more than 2 billion rows). Because it is also not well-optimized to run parallel processing in a cloud-like environment, we are now testing open source database systems, such as Redis\footnote{\url{https://redis.io/}} and GeoMesa\footnote{http://www.geomesa.org/}, to overcome current difficulties.

%%% CALL LIST OF REFERENCES (natbib STYLE) %%%%%%%%%%%%%%%%%%%%%%%%%%


\begin{thebibliography}{}
\bibitem[Albareti et al.(2017)]{alb2017} Albareti, F.~D., Allende Prieto, C., Almeida, A., et al. 2017, ApJS, 233, 25 

\bibitem[Bertin(2006)]{ber2006} Bertin, E. 2006, Automatic Astrometric and Photometric Calibration with SCAMP, ASPC, 351, 112

\bibitem[Bertin \& Arnouts(1996)]{ber1996} Bertin, E. \& Arnouts, S. 1996, SExtractor: Software for source extraction, A\&AS, 117, 393

\bibitem[Berthier et al.(2006)]{berth2006} Berthier, J., Vachier, F., Thuillot, W., et al. 2006, Astronomical Data Analysis Software and Systems XV, 351, 367

\bibitem[Bowell et al.(1995)]{bow1995} Bowell, E., Koehn, B. W., Howell, S. B. et al. 1995, The Lowell Observatory Near-Earth-Object Search: A Progress Report, DPS, 27, 1057

\bibitem[Chang et al.(2015)]{cha2015} Chang, S.-W., Byun, Y.-I., \& Hartman, J. D. 2015, A New Method For Robust High-Precision Time-Series Photometry From Well-Sampled Images: Application to Archival MMT/Megacam Observations of the Open Cluster M37, AJ, 149, 135

\bibitem[Chou et al.(2011)]{chou2011} Chou, J., Howison, M., Austin, B., et al. 2011, Parallel index and query for large scale data analysis. In Proceedings of 2011 International Conference for High Performance Computing, Networking, Storage and Analysis (SC11). ACM, New York, NY, USA, 30, 11

\bibitem[Covey et al.(2007)]{cov2007} Covey, K.~R., Ivezi{\'c}, {\v Z}., Schlegel, D., et al. 2007, Stellar SEDs from 0.3 to 2.5 $\mu$m: Tracing the Stellar Locus and Searching for Color Outliers in the SDSS and 2MASS, AJ, 134, 2398 

\bibitem[Drake et al.(2009)]{dra2009} Drake, A.J., Djorgovski, S. G., Mahabal, A. et al. 2009, First Results from the Catalina Real-time Transient Survey, ApJ, 696, 870

\bibitem[Drake et al.(2014)]{dra2014} Drake, A. J., Graham, M. J., Djorgovski, S. G. et al. 2014, The Catalina Surveys Periodic Variable Star Catalog, ApJS, 213, 9

\bibitem[Flewelling et al.(2016)]{fle16} Flewelling, H.~A., Magnier, E.~A., Chambers, K.~C., et al. 2016, arXiv:1612.05243 

\bibitem[Ivezi{\'c} et al.(2007)]{ive2007} Ivezi{\'c}, {\v Z}., Smith, J. A., Miknaitis, G., et al. 2007, AJ, 134, 973 

\bibitem[Hartman \& Bakos(2016)]{har16} Hartman, J. D. \& Bakos, G. A. 2016, VARTOOLS: A program for analyzing astronomical time-series data, A\&C, 17, 1

\bibitem[Heinze et al.(2018)]{hei18} Heinze, A.~N., Tonry, J.~L., Denneau, L., et al. 2018, arXiv:1804.02132 

\bibitem[Kaiser et al.(2002)]{kai2002} Kaiser, N., Aussel, H., Burke, B.~E., et al. 2002, SPIE, 4836, 154 

\bibitem[Kim et al.(2009)]{kim2009} Kim, D.-W., Protopapas, P., Alcock, et al., 2009, Detrending time series for astronomical variability surveys, MNRAS, 397, 558

\bibitem[Kim et al.(2016a)]{kim2016a} Kim, S.-L., Lee, C.-U., Park, B.-G., et al. 2016a, KMTNet: A Network of 1.6 m Wide-Field Optical Telescopes Installed at Three Southern Observatories, JKAS, 49, 37

\bibitem[Kim et al.(2016b)]{kim2016b} Kim, S.-L., Cha, S.-M., Lee, C.-U., et al. 2016b, Crosstalk Correction of the KMTNet Mosaic CCD Image, PKAS, 31, 35

\bibitem[Lagerkvist et al.(1989)]{lag1989} Lagerkvist, C.-I., Harris, A.~W., \& Zappala, V. 1989, Asteroid lightcurve parameters, Asteroids II, 1162

\bibitem[Larson et al.(2003)]{lar2003} 	Larson, S., Beshore, E., Hill, R et al. 2003, The CSS and SSS NEO surveys, DPS, 35, 3604

\bibitem[Lin et al.(2014)]{lin14} Lin, C.-H., Ip, W.-H., Lin, Z.-Y., Yoshida, F., \& Cheng, Y.-C. 2014, Research in Astronomy and Astrophysics, 14, 311

\bibitem[Liu et al.(2014)]{liu2014} Liu, Y.-b., Wang, F., Ji, K.-f. et al. 2014, NVST Data Archiving System Based On FastBit NoSQL Database, JKAS, 47, 115

\bibitem[Miceli et al.(2008)]{mic2008} Miceli, A., Rest, A., Stubbs, C. W. et al. 2008, Evidence for Distinct Components of the Galactic Stellar Halo from 838 RR Lyrae Stars Discovered in the LONEOS-I Survey, ApJ, 678, 865

\bibitem[Moon et al.(2016)]{moon2016} Moon, H.-K., Kim, M.-J., Yim, H.-S., et al. 2016, Asteroids: New Observations, New Models, 318, 306 

\bibitem[Mueller et al.(2011)]{mue11} Mueller, M., Delbo', M., Hora, J.~L., et al. 2011, AJ, 141, 109 

\bibitem[Palaversa et al.(2013)]{pal2013} Palaversa, L., Ivezi{\'c}, {\v Z}., Eyer, L. et al. 2013, Exploring the Variable Sky with LINEAR. III. Classification of Periodic Light Curves, AJ, 146, 101

\bibitem[Peacock(1983)]{pea1983} Peacock, J. A. 1983, MNRAS, 202, 615 

\bibitem[Press et al.(1992)]{pre1992} Press, W. H., Teukolsky, S. A., Vetterling, W. T., \& Flannery, B. P. 1992, Numerical Recipe (Cambridge: Cambridge Univ. Press)

\bibitem[Rebassa-Mansergas et al.(2013)]{reb13} Rebassa-Mansergas, A., Agurto-Gangas, C., Schreiber, M.~R., G{\"a}nsicke, B.~T., \& Koester, D.\ 2013, MNRAS, 433, 3398 

\bibitem[Ruan et al.(2012)]{rua2012} Ruan, J. J., Anderson, S. F., MacLeod, C. L. et al. 2012, Characterizing the Optical Variability of Bright Blazars: Variability-based Selection of Fermi Active Galactic Nuclei, ApJ, 760, 51

\bibitem[Sesar et al.(2011)]{ses2011} Sesar, B., Stuart, J. S., Ivezi{\'c}, {\v Z}. et al. 2011, Exploring the Variable Sky with LINEAR. I. Photometric Recalibration with the Sloan Digital Sky Survey, AJ, 142, 190

\bibitem[Sesar et al.(2013)]{ses2013} 	Sesar, B., Ivezi{\'c}, {\v Z}., Stuart, J. S. et al. 2013, Exploring the Variable Sky with LINEAR. II. Halo Structure and Substructure Traced by RR Lyrae Stars to 30 kpc, AJ, 146, 21

\bibitem[Shin \& Byun(2004)]{shin2004} Shin, M.-S., \& Byun, Y.-I. 2004, Efficient Period Search for Time Series Photometry, JKAS, 37, 79 

\bibitem[Shin et al.(2009)]{shin2009} Shin, M.-S., Sekora, M., \& Byun, Y.-I. 2009, Detecting variability in massive astronomical time series data - I. Application of an infinite Gaussian mixture model, MNRAS, 400, 1897 

\bibitem[Shin et al.(2012)]{shin2012} Shin, M.-S., Yi, H., Kim, D.-W. et al. 2012, Detecting Variability in Massive Astronomical Time-series Data. II. Variable Candidates in the Northern Sky Variability Survey, AJ, 143, 65 

\bibitem[Skrutskie et al.(2006)]{skr2006} Skrutskie, M. F., Cutri, R. M., Stiening, R. et al. 2006, The Two Micron All Sky Survey (2MASS), AJ, 131, 1163

\bibitem[Stokes et al.(2000)]{sto2000} Stokes, G. H., Evans, J. B., Viggh, H. E. M., et al. 2000, Lincoln Near-Earth Asteroid Program (LINEAR), Icar, 148, 21

\bibitem[Torrealba et al.(2015)]{tor2015} Torrealba, G., Catelan, M., Drake, A. J. et al. 2015, Discovery of $\sim$9000 new RR Lyrae in the southern Catalina surveys, MNRAS, 446, 2251

\bibitem[van Dokkum(2001)]{van2001} van Dokkum, P. G. 2001, Cosmic-Ray Rejection by Laplacian Edge Detection, PASP, 113, 1420

\bibitem[Watson et al.(2017)]{wat2017} Watson, C., Henden, A.~A., \& Price, A. 2017, VizieR Online Data Catalog: AAVSO International Variable Star Index VSX, 1,  

\bibitem[Wolf et al.(2018)]{wol2018} Wolf, C., Onken, C. A., Luvaul, L. C. et al. 2018, SkyMapper Southern Survey: First Data Release (DR1), Arxiv:1801.07834

\bibitem[Wu et al.(2009)]{wu2009} Wu, K., Ahern S., Bethel, E. W. et al. 2009, FastBit: interactively searching massive data, JPhCS, 180, 012053

\bibitem[Yim et al.(2016)]{yim2016} Yim, H.-S., Kim, M.-J., Bae, Y.-H., et al. 2016, Asteroids: New Observations, New Models, 318, 311 

\end{thebibliography}
\end{document}